\documentclass[pre,superscriptaddress,tightenlines,twocolumn,showpacs]{revtex4-1}
\pdfoutput=1

\usepackage{latexsym,amsmath,amsfonts,tabularx,amssymb,bm,empheq}
\usepackage{subfigure}

\usepackage{color}
\usepackage[dvipsnames]{xcolor}
\definecolor{nblue}{RGB}{28,130,185}

\definecolor{cgreen}{RGB}{76,153,0}

\definecolor{myorange}{RGB}{245,156,74}

\usepackage{hyperref}
\hypersetup{
  colorlinks=true,
  citecolor=magenta,
  urlcolor=-myorange
}

\newcommand{\bea}{\begin{eqnarray}}
\newcommand{\eea}{\end{eqnarray}}

\newcommand\subrel[2]{\mathrel{\mathop{#2}\limits_{#1}}}

\newcommand{\dfonc}[2]{\frac{\delta#1}{\delta#2}}
\newcommand{\dd}{\mathrm{d}}
\newcommand{\ddd}{\mathcal{D}}
\newcommand{\ee}{\mathrm{e}}

\newcommand{\bra}{\left\langle}
\newcommand{\ket}{\right\rangle}

\def\simge{\mathrel{%
   \rlap{\raise 0.511ex \hbox{$>$}}{\lower 0.511ex \hbox{$\sim$}}}}
\def\simle{\mathrel{
   \rlap{\raise 0.511ex \hbox{$<$}}{\lower 0.511ex \hbox{$\sim$}}}}

\def\simle{\mathrel{
   \rlap{\raise 0.511ex \hbox{$<$}}{\lower 0.511ex \hbox{$\sim$}}}}

\def\simge{\mathrel{%
    \rlap{\raise 0.511ex \hbox{$>$}}{\lower 0.511ex \hbox{$\sim$}}}}

\begin{document}

\title{Nonuniversality in the erosion of tilted landscapes}

\author{Charlie Duclut and Bertrand Delamotte}
\affiliation{
Laboratoire de Physique Th\'eorique de la Mati\`ere Condens\'ee, UPMC,
CNRS UMR 7600, Sorbonne Universit\'es, 4, place Jussieu, 75252 Paris Cedex 05, France
}

\date{\today}

\begin{abstract}
The anisotropic model for landscapes erosion proposed by Pastor-Satorras  and Rothman in~[R.~Pastor-Satorras and D.~H.~Rothman, \href {\doibase 10.1103/PhysRevLett.80.4349}{Phys. Rev. Lett. \textbf{80}, 4349 (1998)}] is believed to capture the physics of erosion at intermediate length scale ($\lesssim3$~km), and to account for the large value of the roughness exponent $\alpha$ observed in real data at this scale. Our study of this model -- conducted using the nonperturbative renormalization group (NPRG) -- concludes on the nonuniversality of this exponent because of the existence of a line of fixed points. Thus the roughness exponent depends (weakly) on the details of the soil and the erosion mechanisms. We conjecture that this feature, while preserving the generic scaling  observed in real data, could explain the wide spectrum of values of $\alpha$ measured for natural landscapes.
\end{abstract}

\maketitle

\section{Introduction}

Landscapes are known to exhibit scale invariance~\cite{dodds2000}, and Mandelbrot even considered the stretch of a coastline to introduce his notion of fractal dimension~\cite{mandelbrot1982}: seen from far away, the coast displays bays and peninsulas, and reveals more and more sub-bays and sub-peninsulas as one looks closer and closer at it. The self-similarity of branching rivers networks -- where brooks merge into creeks that become streams flowing to form rivers -- is also a well-known fact in geomorphology, and led to several phenomenological scaling laws~\cite{horton1945,rodriguez-iturbe2001,somfai1997}.

Our interest in the following will be erosional landscapes such as mountain ranges, that are also scale invariant~\cite{newman1990}. This scale invariance is made obvious when one studies the roughness of a surface, given by the height-height correlation function:
\begin{align}
    C(r) = \sqrt{\bra |h(x+r)- h(x)|^2 \ket} \, ,
\end{align}
where $\bra \cdot \ket$ denotes a spatial averaging (over $x$). This correlation function is shown in various empirical measurements to scale as $C(r) \sim |r|^\alpha$, where $\alpha$ is known as the roughness exponent. 
Although the scaling behaviour of erosional landscapes is a well-documented fact (from field measurements~\cite{mark1984,czirok1993,norton1989,ouchi1992,matsushita1989,chase1992}, laboratory experiments~\cite{hasbargen2000,paola2009} or numerical simulations~\cite{kim2000,caldarelli1997}), an unambiguous and unique value of the roughness exponent $\alpha$ remains elusive. In fact, from the large amount of experimental data available, two features can be extracted: (i) the roughness exponent has a large variability, and it seems to span the whole range between $\alpha \simeq 0.2$ and $\alpha \simeq 1$, and (ii) there is a tendency to find larger values of the roughness exponent ($0.70 \lesssim \alpha \lesssim 0.95$) at intermediate length scales ($\lesssim3$~km), and smaller values ($0.20 \lesssim \alpha \lesssim 0.60$) at larger length scales~\cite{mark1984,chase1992,kalda2003}.

Because of the complexity and variety of the erosion mechanisms (rainfalls and storms, freezing events and changes in temperature, chemical erosion, landslides and avalanches, etc.~\cite{kukal1990}), a model stemming from these mechanisms is out of reach. However, the scale invariance displayed by these systems suggests that the intermediate and large scale physics of these systems is, at least to a large extent, independent of the smallest  scale details, and that a simple phenomenological model that would capture the relevant elements could be sufficient to reproduce this power-law behaviour and predict the value of the roughness exponent.

So far, some necessary elements for this self-similarity to emerge have already been identified~\cite{sornette1993}. First, the flowing of eroded material by diffusion of the soil has of course to be taken into account. In some simple cases such as river deltas, diffusion in itself can be sufficient to explain the delta front profile~\cite{kenyon1985}. However, the nontrivial scaling property of the correlation function $C(r)$ in eroding landscapes is not reproduced with this sole ingredient. Then, a phenomenological noise term taking into account most of the underlying stochastic phenomena contributing to the erosion must be included~\cite{sornette1993,pelletier2007}. Combining diffusion and noise, one gets the Edwards-Wilkinson noisy diffusion equation. Although this equation yields scale invariance in $d=1$ with a nonvanishing value of $\alpha$, this property is lost in larger dimensions since $\alpha=0$ in this model for all $d>1$~\cite{edwards1982}. Finally, some nonlinearity in the model is mandatory to explain the occurrence of nontrivial values of $\alpha$~\cite{newman1990,roering1999}. The combination of these three elements is minimal to get scaling features in an erosive model.

Amongst the equations displaying the features highlighted above, the Kardar-Parisi-Zhang (KPZ) equation stands out of the crowd~\cite{kardar1986}. First derived and famous in the context of surface growth, the KPZ equation is also thought to describe isotropic erosion of landscapes~\cite{sornette1993}, and predicts $\alpha \simeq 0.4$ in $d=2$~\cite{kloss2012,kelling2011}.

However, although a description of erosion by the KPZ equation seems satisfactory for large scale landscapes, where erosion is indeed isotropic and where the KPZ prediction for the roughness exponent $\alpha$ seems to meet the experimental data (for which $0.20 \lesssim \alpha \lesssim 0.60$), it is not the case for intermediate length scales, where erosion occurs along a preferred direction (the slope of the mountain), and the KPZ equation -- which is isotropic -- fails to capture this important additional ingredient and underestimate the roughness exponent (which is of order $0.70 \lesssim \alpha \lesssim 0.95$)~\cite{dodds2000}. In addition, the KPZ equation is also nonconservative, a feature that is not realistic for smaller scale erosion~\cite{pelletier2007}.

To bridge this gap, Pastor-Satorras and Rothman suggested a nonlinear yet conservative description, and to add anisotropy on top of the three main ingredients discussed above~\cite{pastor-satorras1998,pastor-satorras1998a}. Their perturbative Renormalization Group (RG) analysis that retains only one coupling constant yields exponents in surprisingly good agreement with field measurements. Unfortunately, a recent paper by Antonov and Kakin~\cite{antonov2017a} revealed a mistake in their analysis, showing that there is not a single but infinitely many relevant coupling constants in the theory, which invalidates their results. Antonov and Kakin are however unable to predict the value of the roughness exponent~$\alpha$, but they suggest that the correct model has a line of fixed points, and therefore possibly a continuous range of values for $\alpha$ if this line is attractive, which they cannot show. Moreover, Antonov and Kakin's paper is focused only on a single type of noise (the isotropic noise, which we describe in more details in the following), while Pastor-Satorras and Rothman studied in addition a more interesting model involving a static noise. In this second model, it is not known whether a line of fixed points also exists.

In this paper, we tackle the anisotropic erosion model with  two different kinds of noise using the nonperturbative RG (NPRG)~\cite{berges2002} (for an introduction, see~\cite{gies2012,delamotte2012}), which is perfectly suited for studying a model involving infinitely many coupling constants, since the NPRG is functional in essence. We do agree with Antonov and Kakin about the infinite number of coupling constants involved in the model and with the fact that any truncation retaining only a finite number of them yields wrong predictions in the case of the isotropic noise. We show in addition that this conclusion holds for the two types of noise. 

Furthermore, we are able to integrate numerically the flow equation, and find that in the case of the static noise, there indeed exists for this model an interval  of stable fixed points in the case of physical interest $d=2$. This interval shrinks to a single fixed point, the trivial Edwards-Wilkinson fixed point, in the case of an isotropic noise. This result is of course in marked disagreement with those of~\cite{pastor-satorras1998,pastor-satorras1998a} and in partial disagreement with those of~\cite{antonov2017a} in which it is argued that the isotropic noise case could yield nontrivial exponents.

Moreover, although we are not able to predict whether the whole line of fixed points can be reached from realistic initial conditions, the very existence of this line of fixed points could be a first step to explain the large variability observed in the experimental values of the roughness exponent $\alpha$. Let us also emphasize that despite the very simple formulation of this erosion model, its RG equation displays very interesting features that we describe in the following.

\section{Anisotropic erosion model}

We briefly recall the main features of the model defined in \cite{pastor-satorras1998a}. Our aim is to describe the erosion -- that is the evolution of the height $h(\vec x,t)$ -- of a surface with a fixed mean tilt (e.g. the slope of a mountain) which introduces an intrinsic anisotropy in the model. This preferred direction is determined by a unit vector that we denote $\vec e_\parallel$. Thus, the $d$-dimensional horizontal position $\vec x$ can be decomposed as $\vec x = \vec x_\bot + x_\parallel \vec e_\parallel$ with $\vec x_\bot \cdot \vec e_\parallel = 0$, and $\vec x_\bot$ is therefore a $(d-1)$-dimensional vector. We also define the derivative in the slope direction as $\partial_\parallel\equiv \partial/\partial x_\parallel$ and in the transverse direction as $\nabla_\bot \equiv (\partial/ \partial x_{\bot,i})$ with $i=1 \dots (d-1)$.

 The equation derived by Pastor-Satorras and Rothman in \cite{pastor-satorras1998,pastor-satorras1998a} to describe the evolution of the height profile is a minimal Langevin equation that takes into account diffusion, nonlinearity, noise, and anisotropy. It reads:
\begin{align}
\partial_t h (\bm x)\! = \! \nu_\parallel \partial_\parallel^2 h(\bm x) +\! \nu_\bot \nabla_\bot^2 h(\bm x) +\! \partial_\parallel^2 B(h(\bm x)) +\! \xi(\bm x)
\label{eq_Langevin}
\end{align}
where $\bm x \equiv (\vec x,t)$, the function $B(h)$ is an odd function of the height $h$ and represents the non-linearity, and  $\xi$ is a stochastic noise. As usual, the above Langevin equation has to be understood in the It\=o sense. The non-linear function $B(h)$ takes into account the fact that the flow of water carrying the soil -- and thus responsible for the erosion -- increases with the slope, and is therefore stronger in the downhill direction $\vec e_\parallel$.
The noise probability distribution $P(\xi)$ is
\begin{align}
	P(\xi ) \propto \ee^{-\frac{1}{4D}\int _{\bm x,t'} W(t-t') \xi (x,t) \xi (x,t') }
\end{align}
with $\int _{\bm x} \equiv \int \dd^d x \,\dd t$ (notice that we now drop the arrow above the spatial vector $\vec x$ to alleviate the notation), and the noise correlations are:
\begin{align}
	\bra \xi (\bm x) \xi (\bm{x'}) \ket = 2D\, W(t-t') \delta  ^d (x-x') \, ,
\end{align}
where $W(t-t')=1$ for a static noise, and $W(t-t')=\delta (t-t')$ for an isotropic noise. In this model, the choice of the noise is paramount~\cite{pelletier2007}, since different noises will lead to different universality classes~\cite{caldarelli1997}, different critical dimensions, and therefore either to a trivial ($\alpha=0$), or non-trivial roughness exponent in $d=2$ as we will see in the following. A static (or quenched) noise $W(t-t')=1$ expresses the fact that different types of soil (with various erodibility) can be originally present, whereas an thermal (or isotropic) noise $W(t-t')=\delta (t-t')$ is more suited for mimicking the action of rainfalls over the eroding land. As will be shown in the following, the former leads to a nontrivial roughness exponent in $d=2$, whereas the latter results in smooth landscapes.

From the Langevin equation (\ref{eq_Langevin}), an equivalent field theory can be derived using the Martin--Siggia--Rose--de Dominicis--Janssen (MSRDJ) approach \cite{martin1973,janssen1976,dedominicis1976}. In this formalism, the mean value (over the different realizations of the noise) of a given observable $\mathcal{O}[h]$ is given by:
\begin{align}
	\bra \mathcal{O}[h] \ket _{\xi} &=\int \ddd h  \ddd \tilde{h} \,	\ee ^{-\mathcal{S}[h,\tilde{h}]} \mathcal{O}[h] \label{eq_sansJacob}
\end{align}
with the action
\begin{align}
\begin{split}
    \mathcal{S}[h,\tilde{h}] = \int _{\bm x} \tilde{h}  &
        \left( \partial_t h - \nu_\parallel \partial_\parallel^2 h - \nu_\bot \nabla_\bot^2 h - \partial_\parallel^2 B(h) \right) \\
        & - \int_{x,t,t'} W(t-t') \tilde{h}(x,t) \tilde h(x,t')
         \, .
        \label{eq_action}
\end{split}
\end{align}
Notice that within this formalism the functional integral over $\tilde{h}$ (which is called the ``response'' field) is performed along the imaginary axis, whereas $h$ is a real field. Notice also that up to a rescaling of the time $t$, the longitudinal direction $x_\parallel$, and of the fields $\tilde h$ and $h$, one can set $\nu_\parallel = \nu_\bot= D =1$, which is the normalization we keep in the following and which simplifies the symmetry analysis.

\section{Nonperturbative RG}

In this section we describe briefly the implementation of the nonperturbative RG (NPRG) formalism in the context of a nonequilibrium model~\cite{canet2007,canet2011a}. As in equilibrium statistical physics, the starting point of the field theory is the analog of the partition function associated with the previous action $\mathcal S$ defined in Eq.~(\ref{eq_action}), and which reads:
\begin{align}
    \mathcal{Z}[j,\tilde{j}] = \int \mathcal{D} h  \mathcal{D}\tilde{h} \, 
    \mathrm{e}^{-\mathcal{S}+\int_{\bm x}J(\bm x)^T \cdot H (\bm x)}
\end{align}
where we use a matrix notation and define the following vectors
\begin{align}
    H (\bm x)= 
        \left( 
             \begin{array}{c} h(\bm x) \\ \tilde{h}(\bm x)  \end{array} 
        \right) 
    \quad \text{and} \quad
    J(\bm x)= 
        \left( 
             \begin{array}{c} j(\bm x) \\ \tilde{j}(\bm x)  \end{array} 
        \right) \, .
\end{align}
As in equilibrium, the generating functional of the connected correlation and response functions is  $\mathcal{W}[J] = \log \mathcal{Z}[J]$. We also introduce its Legendre transform, the generating functional of the one-particle irreducible correlation functions $\Gamma[\Phi]$, where $\Phi= \bra H \ket$.

In order to determine the effective action $\Gamma$, we apply the NPRG formalism and write a functional differential equation which interpolates between the microscopic action~$\mathcal{S}$ and the effective action~$\Gamma$. The interpolation is performed through a momentum scale $k$ and by integrating over the fluctuations with momenta $\vert q\vert>k$, while those with momenta $\vert q\vert<k$ are frozen. At scale $k=\Lambda$, where $\Lambda$ is the ultra-violet cutoff imposed by the (inverse) microscopic scale of the model (e.g. the lattice spacing), all fluctuations are frozen and the mean-field approximation becomes exact; at scale $k \to 0$, all the fluctuations are integrated over and the original functional $\mathcal{Z}$ is recovered. The interpolation between these scales is made possible by using a regulator $\mathcal{R}_k(\bm x)$, whose role is to freeze-out all the fluctuations with momenta $\vert q\vert<k$. This regulator is introduced by adding an extra term to the action and thus defining a new partition function $\mathcal Z_k$:
\begin{align}
    \mathcal{Z}_k[j,\tilde{j}] = \int \mathcal{D} h  \mathcal{D}\tilde{h} \, 
    \mathrm{e}^{-\mathcal{S}-\Delta \mathcal S_k+\int_{\bm x}J(\bm x)^T \cdot H (\bm x)}
    \label{eq_partitionFunction}
\end{align}
with
\begin{align}
    \Delta\mathcal{S}_k= \frac{1}{2}\int_{\bm x,\bm{x'}} H(\bm x)^T \cdot \mathcal{R}_k(\bm x-\bm{x'}) \cdot H (\bm{x'})
    \label{eq:def_regulator}
\end{align}
where $\mathcal{R}_k$ is a $2\times 2$ regulator matrix, depending both on space and time, and whose task is to cancel slow-mode fluctuations.
Let us first recall that the MSRDJ formalism together with It\=o's prescription does not allow for a term in the action not proportional to the response field $\tilde h$. This implies that there is no cutoff term in the $h-h$ direction, and  the regulator matrix defined in Eq.~(\ref{eq:def_regulator}) can be written in full generality as
\begin{align}
    \mathcal{R}_k(\bm x)=\left( 
         \begin{array}{cc} 0 & R_{1,k}( x,t) \\ R_{1,k}( x,-t) &  2 R_{2,k}( x,t)  \end{array} 
    \right) \, ,
\end{align}
where the minus sign in $R_{1,k}(x,-t)$ is a consequence of $\Delta \mathcal{S}_k$ being written in a matrix form and the factor 2 in front of $R_{2,k}$ has been included for convenience. 

In the following, we  only consider a space regulator, that is, a regulator which is trivial in the time direction, and we also discard the noise modification $R_{2,k}$ (see \cite{duclut2017} for further discussion of a frequency regulator), such that 
\begin{align}
    \mathcal{R}_k(\bm x)=\left( 
         \begin{array}{cc} 0 & R_k(x) \delta(t) \\ R_k(x) \delta(t) &  0 \end{array} 
    \right) \, .
\end{align}
In this paper we use the $\Theta$-regulator which allows for an analytic computation of the integrals over momentum, and which is defined in Fourier space as
\begin{align}
    R_k(q) &= (k^2-q^2) \Theta (k^2-q^2)
    \label{eq_Litim}
\end{align}
where $\Theta (q)$ is the Heavyside step-function ($\Theta (q<0)=0$ and $\Theta (q\geq0)=1$). Notice that we have kept the same name for the function and its Fourier transform, which is defined as: 
\begin{align}
f(\bm q) \equiv \int_{\bm x} f(\bm x) \, \ee^{-i(q x - \omega t)} \, .
\label{eq_FTconvention}
\end{align}

We also define the effective average action $\Gamma_k$ as a modified Legendre transform  of $\mathcal{W}_k[J] = \log \mathcal{Z}_k[J]$~\cite{wetterich1993}:
\begin{align}
\begin{split}
    &\Gamma_k[\Phi]+\mathcal{W}_k[J] =  \\
    &\int_{\bm x} J^T\cdot \Phi - \frac{1}{2} \int_{\bm x,\bm{x'}} \Phi(\bm x)^T \cdot \mathcal{R}_k(\bm x-\bm{x'}) \cdot \Phi (\bm{x'})
    \label{eq_LegendreTransform}
\end{split}
\end{align}
in such a way that $\Gamma_k$ coincides with the action at the microscopic scale ($\Gamma_{k=\Lambda}=\mathcal S$) and with $\Gamma$ at $k=0$ ($\Gamma_{k=0}=\Gamma$), when all fluctuations have been integrated over. The evolution of the interpolating functional $\Gamma_k$ between these two scales is given by the Wetterich equation~\cite{wetterich1993,morris1994}:
\begin{align}
    \partial_k \Gamma_k [\Phi] &= \frac{1}{2} \text{Tr} \int_{\bm x,\bm{x'}} \partial_k \mathcal{R}_k(\bm x-\bm{x'}) \cdot G_k [\bm x,\bm{x'};\Phi]
    \label{eq_Wetterich}
\end{align}
where $G_k [\bm x,\bm{x'};\Phi] \equiv [ \Gamma_k^{(2)}+\mathcal{R}_k]^{-1}[\bm x,\bm{x'};\Phi]$ is the full, field-dependent, propagator and $\Gamma_k^{(2)}$ is the $2 \times 2$ matrix whose elements are the $\Gamma_{k,ij}^{(2)}$ defined such that:
\begin{align}
    \Gamma_{k,i_1,\cdots,i_n}^{(n)}[{\bm x_i};\Phi] &= \frac{\delta^n \Gamma_k[\Phi]}{\delta \Phi_{i_1}(\bm x_1)\cdots \delta \Phi_{i_n}(\bm x_n)} \, .
\end{align}
The Wetterich equation~(\ref{eq_Wetterich}) represents an exact flow equation for the effective average action $\Gamma_k$, which we solve approximately by restricting its functional form. We use in the following the derivative expansion (DE), stating that instead of following the full $\Gamma_k$ along the flow, only the first terms of its series expansion in space and time derivatives of $\Phi$ are considered. This method is very efficient and has led both at and out of equilibrium to many accurate and original results \cite{canet2003a,canet2005a,kloss2014,*canet2011,*canet2012,delamotte2004,benitez2008,caffarel2001,*holovatch2004,*peles2004,*delamotte2004a,canet2004,*canet2004a,*canet2003,*canet2005,tissier2010,tissier2008,*tissier2012,*tissier2012a}.
The terms retained in this derivative expansion have to be consistent with the symmetries of the action~$\mathcal S$, and we therefore discuss them before giving an explicit ansatz for $\Gamma_k$.

\section{Symmetries}

In order to find a meaningful and simple ansatz for the effective average action $\Gamma_k$, we start by studying the symmetries of the action. We consider the following shift-gauged symmetry:
\begin{align}
    \tilde{h}'(\bm x) = \tilde{h}(\bm x) + \varepsilon (x_\bot,t)
    \label{eq_symmetry}
\end{align}
where $\varepsilon$ is an arbitrary infinitesimal function. The action~(\ref{eq_action}) is not strictly invariant under the transformation~(\ref{eq_symmetry}), but since the variations of the action following this transformation are linear in the fields, it also yields useful Ward identities~\cite{canet2011,canet2016}. Under transformation~(\ref{eq_symmetry}), the integral~(\ref{eq_partitionFunction}) remains unchanged, which yields:
\begin{align}
\begin{split}
    \int_{\bm x} &\left[ \tilde j \varepsilon  - \varepsilon \partial_t \bra h \ket  + \varepsilon \nabla_\bot^2 \bra h \ket - \int_{\bm x'} \varepsilon R_k \bra h \ket \right] \\
    & + 2 \int_{\bm x,t'} W(t-t')\,  \varepsilon(x_\bot,t) \bra \tilde h(x,t') \ket  = 0 \, .
\end{split}
\end{align}
Notice that we have integrated by parts the terms involving a derivation with respect to $x_\parallel$, and that the boundary terms that result from this integration by parts vanish because of the symmetry $x_\parallel \to - x_\parallel$.

Then, using the definition~(\ref{eq_LegendreTransform}) of the modified Legendre transform to eliminate the external field $\tilde j$, and using the fact that, by definition, $\bra h \ket = \phi$ and $\langle \tilde h \rangle = \tilde \phi$, the previous expression becomes:
\begin{align}
    \int_{\bm x} \! \left[ \dfonc{\Gamma_k}{\tilde \phi} \!-\! \partial_t \phi \!+\! \nabla_\bot^2 \phi \!+\!2 \!\! \int_{t'} \!  W(t-t') \tilde \phi(x,t')  \right] \! \varepsilon(x_\bot,t)\! =\! 0\, .
\end{align}
Since this equality is true for any function $\varepsilon(x_\bot,t)$, it means that the Fourier transform [defined in Eq.~(\ref{eq_FTconvention})] of the term inside the brackets vanishes at $q_\parallel=0$. Consequently, at $q_\parallel=0$, the functional
\begin{align}
    \Gamma_k\! - \!\! \int_{\bm q} \!\! \tilde \phi (-\bm q) \! \left[ -i\omega +q_\bot^2 \right] \phi(\bm q) +\!\int_{\bm q}\!\! W(\omega) \tilde \phi(-\bm q) \tilde \phi(\bm q)
\end{align}
vanishes under transformation~(\ref{eq_symmetry}). It finally means that only the terms $\partial_\parallel h$ and $\partial_\parallel B(h)$ [which are invariant under~(\ref{eq_symmetry})] are renormalized, while the terms $\int \tilde \phi \partial_t \phi$, $\int \tilde \phi \nabla_{\bot}^2 \phi$, and $\int W(t-t') \tilde{\phi}(x,t) \tilde \phi(x,t')$ are not. Thus, at lowest order in the space and time derivatives, the most general ansatz for the effective average action $\Gamma_k[\phi,\tilde \phi]$ reads
\begin{align}
\begin{split}
    \Gamma_k[\phi,\tilde \phi] = \int_{x,t} &\tilde \phi(x,t) \left[ 
    \partial_t \phi - \nabla_{\bot}^2 \phi - \partial_{\parallel}^2 A_k(\phi) \right] \\
    &- \int_{x,t,t'} W(t-t') \tilde{\phi}(x,t) \tilde \phi(x,t') \, .
    \label{eq_Gammak}
\end{split}
\end{align}
We conclude that at this order only one function, $A_k(\phi)$, has a nontrivial renormalization flow that we derive in the following.

\section{Upper critical dimension and controversies}

Before deriving the flow equation and giving the results using the NPRG, we discuss here the upper critical dimension of this model, and try to clarify the misunderstanding about the relevance of some operators. First, depending on the nature of the noise, isotropic or static, the model has different upper critical dimensions. This upper critical dimension is $d_c^{\, \text{stat}}=4$ in the case of a static noise, and  $d_c^{\, \text{iso}}=2$ in the case of an isotropic noise, as already stated in~\cite{pastor-satorras1998}. 

The computation of the upper critical dimension is made very simple once the model has been cast into its simplest form~(\ref{eq_Gammak}) using symmetry considerations. From this equation, we find that the engineering dimension of the field $\phi$ (expressed in momentum scale) is:
\begin{align}
[\phi] = \frac{2(d-2\kappa)}{3}
\end{align}
where $\kappa = 1$ for an isotropic noise, and $\kappa=2$ for a static noise. Therefore, a coupling constant in front of a $\phi^i$ term is irrelevant for $d>d_c=2\kappa$, which indeed yields the previous upper critical dimensions. 

However, the important and surprising feature of this model is that, exactly at the upper critical dimension $d=d_c^{\, \text{stat}}$ or $d=d_c^{\, \text{iso}}$, the dimension of the field $\phi$ vanishes, meaning that all terms $\int_{\bm x} \tilde\phi\,\partial_{\parallel}^2\phi^n$ coming from the expansion of the function $A_k(\phi)$ in Eq.~(\ref{eq_Gammak}) are equally relevant, as pointed out in~\cite{antonov2017a,antonov2017} in the isotropic case. It therefore invalidates the whole approach of~\cite{pastor-satorras1998,pastor-satorras1998a} since infinitely many coupling constants were discarded. We indeed show in the following that truncating the function $A_k$ greatly modifies the physics and the computation of the critical exponent of the model.

\section{Flow equation}

We now compute the flow of the function $A_k(\phi)$, which we define as:
\begin{align}
A_k(\phi) = \frac{1}{\Omega} \left. \left( \partial_{p_{\parallel}^2} \text{FT} \left( \dfonc{\Gamma_k}{\tilde \phi (z)} \right) (\bm p) \right) \right|_{\phi(x,t)=\phi,\bm p=0}
\label{eq_Ak}
\end{align}
where $\Omega$ is the volume of the system, and $\text{FT}(f)(\bm q)$ refers to the Fourier Transform of the function $f(\bm x)$ with the convention~(\ref{eq_FTconvention}). Notice that one has to evaluate it at constant field \textit{after} having performed the momentum derivation. This is unusual in the NPRG context, and we therefore give slightly more details of the derivation of the flow in Appendix~\ref{app_flow}. In order to find a fixed point of the RG flow, one has to write the flow equation in terms of dimensionless variables. We define them in the following way:
\begin{subequations}
\begin{empheq}{align}
    \hat{x}_\bot &= k \,x_\bot \\
    \hat t &= k^{2}\,t \\
    \hat{A}(\hat{\phi}) &= \bar{A}_k^{-1} A_k(\phi) \\
    \hat{x}_\parallel &= k^{1+(d-2\kappa)/3} \bar{A}_k^{-2/3} \,x_\parallel \\
    \hat{\phi} &= k^{(4\kappa-2d)/3} \bar{A}_k^{1/3} \,\phi\\
    \hat{\tilde{\phi}} &= k^{2(\kappa-d)/3} \bar{A}_k^{1/3} \,\tilde{\phi} %
\end{empheq}%
\label{eq_adimensionalisation}%
\end{subequations}%
where we define the running coefficient $\bar{A}_k$ such that $\hat A'(\hat \phi=0)\equiv 1$ where the prime means derivation with respect to $\phi$. In the critical regime, this running coefficient is expected to behave as a power law $\bar{A}_k \sim k^{-\eta_A^*}$, and we therefore define a running exponent $\eta_A(k) = - k \partial_k \ln \bar A_k$ such that $\eta_A(k=0)\equiv \eta_A^*$. The roughness exponent $\alpha$ and the anisotropy exponent $\zeta$ correspond respectively to the anomalous dimension of the field $\phi$ and to the anomalous dimension of the longitudinal direction $x_{\parallel}$. They can thus be expressed in terms of the fixed point value of $\eta_A^*$ as
\begin{align}
    \alpha &\equiv (4\kappa-2d-\eta_A^*)/3 \, , \label{eq_alpha}\\
    \zeta &\equiv 1+(d+2\eta_A^*-2\kappa)/3 \, .\label{eq_zeta}
\end{align}

The flow of the function $\hat A(\hat \phi)$ can be split into two parts:
\begin{align}
k\partial_k \hat{A}(\hat{\phi}) = k\partial_k \hat{A}(\hat{\phi})|_{\text{dim}} + k\partial_k \hat{A}(\hat{\phi})|_{\text{dyn}}
\label{eq_flowA}
\end{align}
where the dimensional part of the flow $k\partial_k \hat{A}(\hat{\phi})|_{\text{dim}}$ directly follows from the previous definitions~(\ref{eq_adimensionalisation}) and reads:
\begin{align}
k\partial_k \hat{A}(\hat{\phi})|_{\text{dim}} &= \eta_A \hat{A}(\hat{\phi}) +\frac{2d+\eta_A-4 \kappa}{3} \hat{\phi} \hat{A}'(\hat{\phi}) \, ,
\end{align}
while the dynamical part of the flow is derived in Appendix~\ref{app_flow} and reads:
\begin{align}
\begin{split}
   &k\partial_k \hat{A}(\hat{\phi})|_{\text{dyn}} = \frac{(3\kappa-2) K_d}{2} \times \\ &\int_{y=0}^{\infty} \int_{\theta=0}^\pi \frac{y^{d/2-\kappa}\sin(\theta)^{d-2} r'(y) \hat{A}''(\hat{\phi})}{\left(r(y)+\sin^2\theta+ \hat{A}'(\hat{\phi})\cos^2\theta \right)^{1+\kappa}}
\label{eq_flowAdyn}
\end{split}
\end{align}
where $K_d=(2^{d-1}\pi^{d/2}\Gamma(d/2))^{-1}=S_{d-1}/(2\pi)^d$ with $S_d$ the surface of the $d$-dimensional unit hypersphere. 
Moreover, the definition of the running anomalous dimension $\eta_A(k)$ provides us with the additional equation $k \partial_k \hat{A}'(0) =0$, which yields:
\begin{align}
\begin{split}
    &\eta_A = \kappa -d/2-\frac{3 (3\kappa-2) K_d}{8 \hat{A}'(0)} \times \\
    & \int_{y=0}^{\infty} \int_{\theta=0}^\pi \frac{y^{d/2-\kappa}\sin(\theta)^{d-2} r'(y) \hat{A}'''(0)}{\left(r(y)+\sin^2\theta+ \hat{A}'(0)\cos^2\theta\right)^{1+\kappa}} \, .
    \label{eq_flowEtaA}
\end{split}
\end{align}
Notice that in Eqs.~(\ref{eq_flowA}) and (\ref{eq_flowEtaA}) the dimension $d$, as well as the nature of the noise $\kappa$ are real parameters that can be chosen at will. Starting from the flow equations (\ref{eq_flowA}) to (\ref{eq_flowEtaA}), one can easily retrieve the one-loop perturbative results obtained in \cite{antonov2017a}, and the truncated results of~\cite{pastor-satorras1998,pastor-satorras1998a}; this is explained in Appendix~\ref{app_perturbative}. 

Notice that in the case of static noise, in $d=2$ and with the $\Theta$ regulator~(\ref{eq_Litim}), the flow equation~(\ref{eq_flowA}) can be rewritten in a much simpler form:
\begin{align}
    k\partial_k \hat{A} = \eta_A \hat{A}+ \frac{\eta_A-4}{3}\hat\phi \hat{A}'-\frac{(1+3\hat{A}')\hat{A}''}{4 (\hat{A}')^{3/2}} 
    \label{eq_flowAlitim}
\end{align}
where we have omitted the argument of $\hat A$ and its $k$-dependence for convenience.

    \section{Line of fixed points}

We now study the properties of the flow equation~(\ref{eq_flowA}). Notice that at the fixed point (namely when $\hat{A}(\hat{\phi})=\hat{A}^*(\hat{\phi})$ such that $k\partial_k \hat{A}^*(\hat{\phi})=0$), the flow equation provides us with an iterative scheme for computing the derivatives $\hat{A}^{*(j)}(0)\equiv a_j$ for all $j$. Indeed, at the fixed point and evaluated at $\hat{\phi}=0$, the derivatives of Eq.~(\ref{eq_flowA}) can be rewritten as:
\begin{subequations}%
\begin{empheq}{align}%
    & f_3 (\eta_A^*, a_3) = 0 \\
    & f_5 (\eta_A^*, a_3, a_5) = 0 \\
    & f_7 (\eta_A^*, a_3, a_5, a_7) = 0 \\
    & \hspace{1.3cm} \vdots \nonumber
\end{empheq}%
\end{subequations}%
where the ${f_i}$ are linear functions of their last argument. For instance, for the static noise in $d=2$ and with the $\Theta$~regulator~(\ref{eq_Litim}), the previous equations yield:
\begin{subequations}%
\begin{empheq}{align}%
    & a_3 = \frac{4}{3}(\eta_A^*-1) \\
    & a_5 = \frac{4}{3} (\eta_A^*-1) (5\eta_A^*-7) \label{eq_a5}\\
    & \hspace{1.7cm} \vdots \nonumber
\end{empheq}%
\end{subequations}%
Therefore, provided that the Taylor expansion of $\hat A^*(\hat\phi)$ around $\hat\phi=0$ can be analytically continued on the whole real axis then a line of fixed points parametrized by the values of $\eta_A^*$ exists, as claimed in~\cite{antonov2017a}. On the other hand, notice that a truncation of $\hat A$ at any finite order will \emph{not} yield a line of fixed points. For instance, writing $\hat A=\hat \phi + a_3/3! \, \hat \phi^3$ means that the coefficient $a_5$ vanishes and thus yields $\eta_A^*=1$ or $\eta_A^*=7/5$ according to Eq.~(\ref{eq_a5}). Instead of improving the accuracy of $\eta_A^*$, increasing the rank of the truncation will rather yield more and more (different) fixed points, with some stable and some unstable. The correct picture is therefore only accessible  when the problem is tackled functionally, that is with the full function $\hat A(\hat \phi)$.

Studying numerically these fixed points as well as their stability is non-trivial as we show in the following, but simple physical arguments already allow us some comments: (i)~the line of fixed point is upper bounded in all dimensions because the roughness exponent $\alpha$ is positive, and we therefore deduce from Eq.~(\ref{eq_alpha}) that $\eta_A^* \leq  2 (2\kappa -d)$; (ii) the anisotropy exponent $\zeta$ characterizes the ratio between the roughness exponent in the transverse direction, $\alpha_\bot\equiv \alpha$, and the roughness exponent in the parallel direction $\alpha_\parallel$~\cite{pastor-satorras1998,pastor-satorras1998a}. In our anisotropic model, we expect this ratio to be larger than 1, i.e., $\zeta\geq 1$, which translates for $\eta_A^*$ as (using Eq.~(\ref{eq_zeta})]: $\eta_A^* \geq (2\kappa-d)/2$.

\begin{figure}[t!]
	\centering 
    \subfigure[ ]
    {\includegraphics[width=0.45\textwidth]{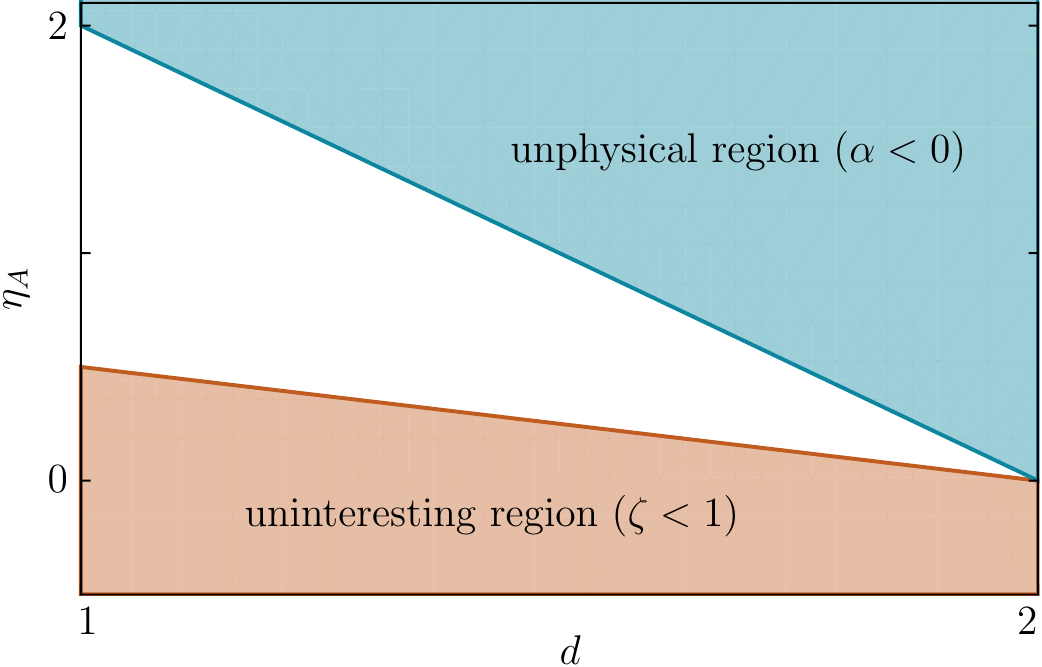}}
    \subfigure[ ]
    {\includegraphics[width=0.45\textwidth]{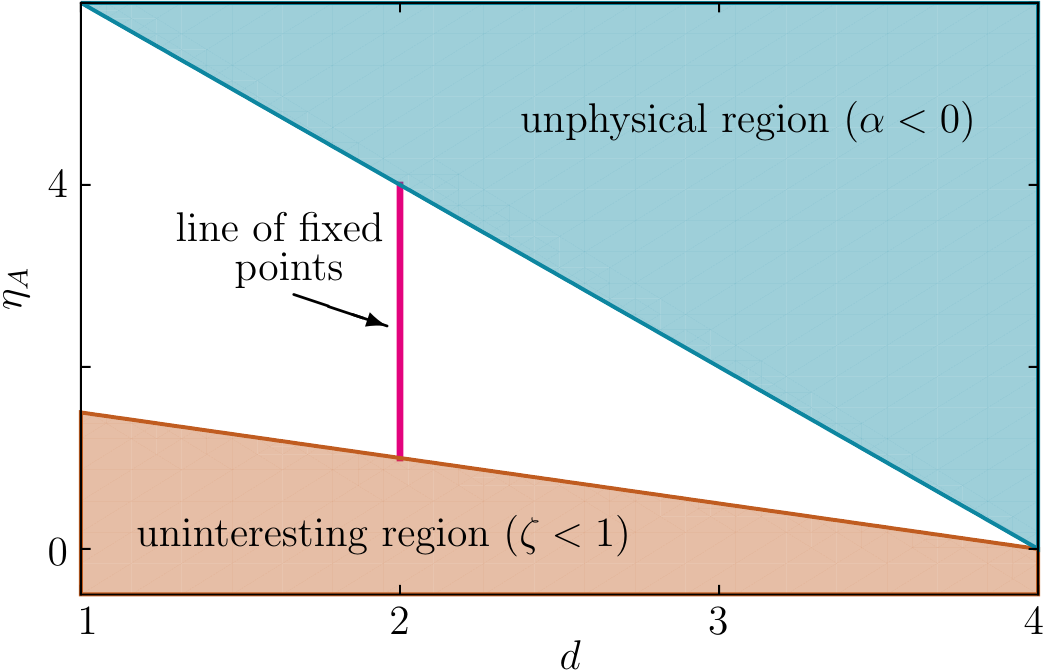}}

    \caption{Critical exponent $\eta_A$ for isotropic (a) and static (b) noises as a function of the physical space dimension $d$. Recall that for landscape erosion, the dimension of interest is $d=2$. The upper colored region is unphysical($\alpha<0$). Its lower boundary is the Edwards-Wilkinson fixed point with $\alpha=0$. The bottom region is the physical yet uninteresting region for which the anisotropy exponent $\zeta$ is lower than 1. In this region, the function behaves like $\hat A_k^*\sim \hat \phi^\gamma$ as $\hat \phi\to\infty$, with $\gamma <1$, and the system does not display the kind of nonlinearity we were looking for. The blank region in between is therefore the interesting region for our model; it ends up in a single point at the upper critical dimension, $d_c^{\,\text{iso}}=2$ (a), or $d_c^{\,\text{stat}}=4$ (b). In the case of the anisotropic noise (b), we see that there is an interval of fixed points (red line) in $d=2$.}
    \label{fig_etaVSd}
\end{figure}

The first inequality is directly encoded in the flow equation since there exists no scaling solution (of the form $\hat A^*(\hat\phi) \sim \hat\phi^\gamma$ at large field) of the fixed point equation~(\ref{eq_flowA}) when $\eta_A^*$ is such that $\alpha<0$. The second inequality also has a signature in the flow equation, more precisely on the scaling form of the fixed point function $\hat A^* (\hat \phi)$: indeed, studying Eq.~(\ref{eq_flowA}) at large field, one finds that the fixed point function should scale as
\begin{align}
\hat{A}^*(\hat{\phi}) \subrel{\hat{\phi}\to \infty}{\sim} \hat{\phi}^{\,\gamma} \quad \text{with} \quad \gamma = \frac{3\eta_A^*}{4\kappa-2d-\eta_A^*}
\label{eq_largePhi}
\end{align}
and the inequality $\zeta\geq 1$ is equivalent to saying that $\hat{A}^*(\hat{\phi})$ is sub-linear at large field, which is not unphysical, but simply does not correspond to the model that we study where we expect non-linearity and a power-law behaviour at large field. These considerations allow us to discard the isotropic noise ($\kappa=1$) since in dimension $d=2=d_c^{\, \text{iso}}$ (the physical dimension of our problem), the only value of $\alpha$ that satisfies both inequalities is the trivial Edwards-Wilkinson exponent $\alpha=0$. Within this erosion model, an isotropic noise can therefore not explain the observed landscapes roughness; see Fig.~\ref{fig_etaVSd}.

    \section{Numerical solution}
    
\begin{figure}[t!]
	\centering 
    \subfigure[ \label{fig_plateaus1a} ]
    {\includegraphics[width=0.45\textwidth]{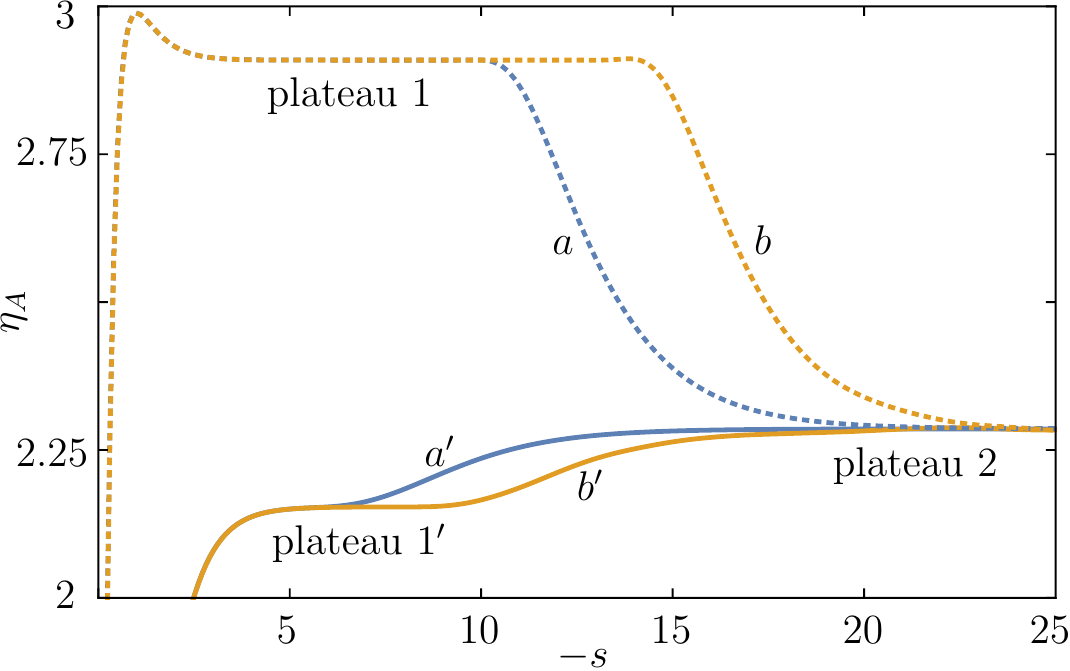}}
    \subfigure[ \label{fig_plateaus1b} ]
    {\includegraphics[width=0.45\textwidth]{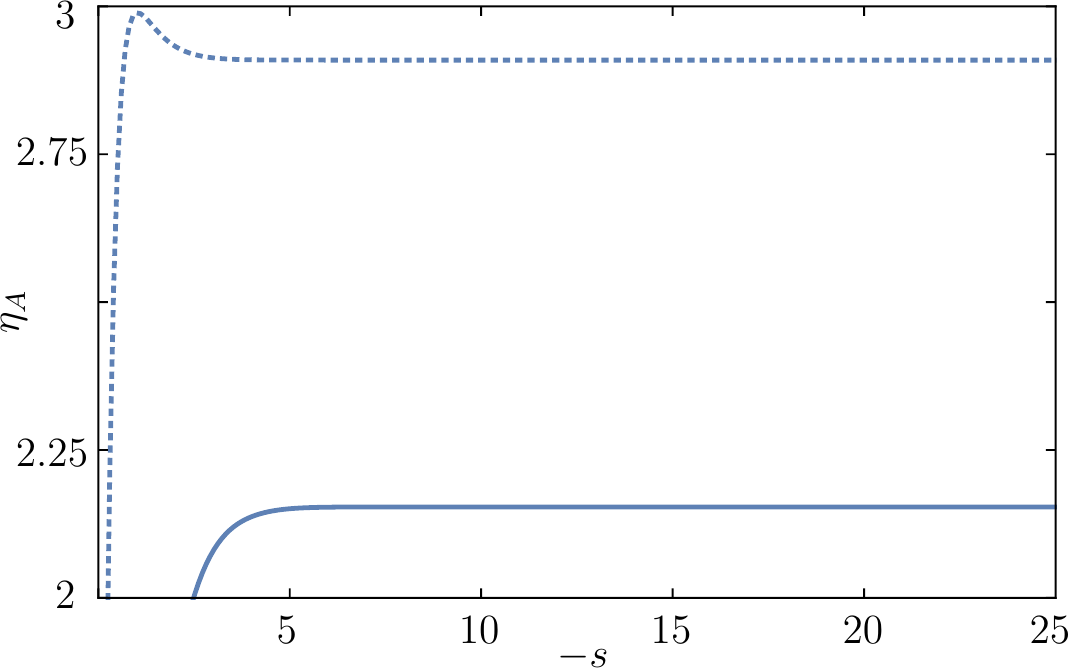}}

    \caption{RG flows ($s=\log (k/\Lambda)$ is the RG time) of the exponent $\eta_A$ for two different initial conditions, obtained by integrating numerically the flow equation~(\ref{eq_flowAlitim}). \textbf{(a)} Dotted lines $a$ and $b$: initial condition with a large field behaviour $\hat A^{\text{init}}(\hat\phi) \sim \hat\phi^{8}$ for which we expect from Eq.~(\ref{eq_largePhi}) an exponent $\eta_A^* \simeq 2.91$ which is indeed what is observed on the plateau 1. Solid lines $a'$ and $b'$: same as above with
    $\hat A^{\text{init}}(\hat\phi) \sim \hat\phi^{3.5}$ and $\eta_A^*\simeq2.15$ which is observed on the plateau 1'. At large $s$, both flows end on the plateau 2. The curves $b$ and $b'$ are obtained by increasing the size of the box $\hat \phi_{\text{max}}$, which increases the length of the plateaus 1 and 1'. \textbf{(b)} Same initial conditions as for (a), but with improved computation of the derivatives of $\hat A$ around  $\hat \phi_{\text{max}}$ (see main text). With this method, the first plateaus 1 and 1' are never left showing that the crossover to plateau 2 is a numerical artifact.}
    \label{fig_plateaus1}
\end{figure}

\begin{figure}[t!]
	\centering 
    {\includegraphics[width=0.45\textwidth]{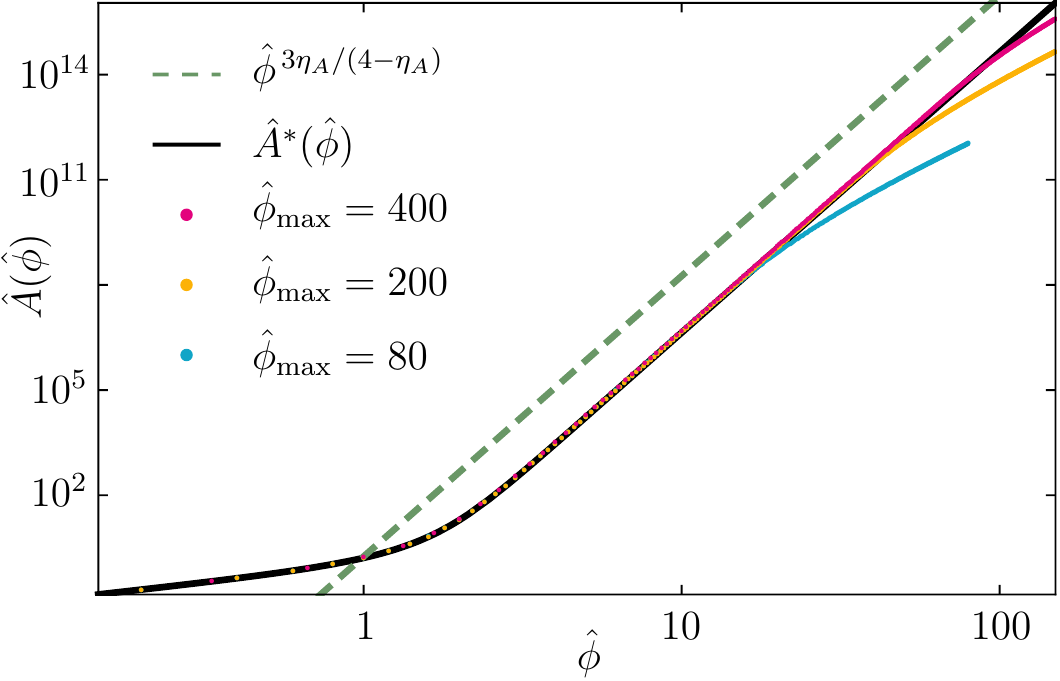}}

    \caption{Solid line: Fixed point solution $\hat A^*(\hat\phi)$ of Eq.~(\ref{eq_flowA}) for $\eta_A=\eta_A^{\text{plateau}}\simeq2.91$. Dashed line: asymptotic behaviour in $\hat \phi^{\, 3\eta_A/(4-\eta_A)}$ with $\eta_A=\eta_A^{\text{plateau}}$. Dots: plateau solution $\hat A^{\text{plateau}}(\hat\phi)$ for $\hat \phi_{\text{max}}=80$ (blue), 200 (yellow), and 400 (red) taken from the numerical solution of Eq.~(\ref{eq_flowA}) at RG time $s=-5$. The plateau solution converges towards the true fixed point solution as $\hat \phi_{\text{max}}$ is increased.}
    \label{fig_Aplateau}
\end{figure}

We are now interested in confirming the existence of the line of fixed points found above from a Taylor expansion around $\hat\phi=0$. We now focus on the case of static noise, in $d=2$ and with the $\Theta$-regulator~(\ref{eq_Litim}), although the method we present remains true for a different noise, dimension or regulator. The flow equation in this case is given by Eq.~(\ref{eq_flowAlitim}). 

We thus solve numerically the fixed point equation: 
$k\partial_k \hat{A}^*(\hat{\phi}) =0 $ 
together with the two boundary conditions $\hat A^*(0)=0$ coming from the fact that $A(\phi)$ is odd and $\hat A^*{}'(0)=1$ which defines $\eta_A(k)$. The numerical integration is performed on a finite grid $\hat \phi \in [0,\hat \phi_{\text{max}}]$.
The derivatives of $\hat A^*$ are then computed on this grid using the usual ``five-point stencil'' method. At the leftmost part of the grid ($\hat \phi=0$), we use the fact that $\hat A^*(-\hat\phi)=-\hat A^*(\hat\phi)$. On the rightmost part of the grid, we do not impose any boundary condition and the derivatives are computed using only points inside the grid. This simple scheme confirms the existence of a line of fixed points: for any given $\eta_A^*$ (such that $\alpha\geq0$) we find a fixed point function $\hat A^*$ solution of Eq.~(\ref{eq_flowAlitim}). The precision of each of these solutions is refined when the size of the box $\hat \phi_{\text{max}}$ or the number of discretization points is increased. In particular, the scaling at large field, Eq.~(\ref{eq_largePhi}), is very well reproduced (at least when $\hat \phi_{\text{max}}$ is large enough) which confirms the global existence of the fixed points. Notice that an exact solution of the fixed point equation (\ref{eq_flowAlitim}) for $\eta_A^*=0$ is available (see Appendix~\ref{app_fixedPoint}) which allows for a check of our numerical solution in this particular case.

The stability of these fixed points is a subtler issue. Usually, the stability analysis is simply performed by linearizing the flow around the fixed point, that is, by computing the (discretized) stability matrix and evaluating its eigenvalues. The sign of these eigenvalues then provides the stability of each fixed point. An alternative path consists in perturbing the fixed point solution: $\hat A(\hat \phi) = \hat A^*(\hat \phi) + \varepsilon \ee^{\lambda s} g(\hat \phi)$ (where $s=\log(k/\Lambda)$ is the RG time) and then solving the differential equation for $g$ while using a shooting method to find the eigenvalues $\lambda$~\cite{bender1978,bervillier2008,bervillier2008a}. In this model however, none of these methods yield reliable results since we do not observe the convergence of the eigenvalues when the size of the box or the number of discretization points is increased.

To tackle this issue, we perform a numerical integration of the flow equation~(\ref{eq_flowAlitim}) starting with different initial conditions $\hat A^{\text{init}}$. We use a Runge-Kutta scheme and the same discretization for the field $\hat \phi$ as explained above for the fixed point equation. For various initial conditions, we observe that $\eta_A(s)$ reaches a first plateau [see Fig.~\ref{fig_plateaus1a}] which is left after a finite RG time. The flows then reach a second plateau where they stay forever. Whereas the position of the first plateaus depends on the initial condition, the second plateau is the same for all initial conditions; this seems to indicate the existence of a unique fully attractive fixed point, for which $\eta_A^*\simeq 2.29$, whereas all the other fixed points are unstable. However, increasing the size of the box $\hat \phi_{\text{max}}$ increases the length of the first plateaus (see Fig.~\ref{fig_plateaus1a}] and it seems that except for numerical stability issues, we could virtually extend these plateaus for an arbitrary long RG time by increasing $\hat\phi_{\rm max}$. We  notice that all the plateau functions $\hat A^{\text{plateau}}(\hat\phi)$ match with the fixed point solutions found by integrating Eq.~(\ref{eq_flowAlitim}) directly at the fixed point and for $\eta_A=\eta_A^{\text{plateau}}$; see Fig.~\ref{fig_Aplateau}. This indicates that the first plateaus correspond to  fixed points that are (numerically) unstable.

To cure the sensitive dependence of the numerical flow on the box size $\hat\phi_{\rm max}$, we proceed to a compactification of the field $\hat \phi$ and define $y=\hat \phi^2/(m+\hat\phi^2)$, with $m$ a free parameter. The whole interval $\hat \phi\in [0,\infty[$ is mapped onto $y\in [0,1[$ that can be discretized. We also compactify the function $\hat A$ and obtain a new function $D(y)$ which remains finite when $y\in[0,1[$. In this compactified version, the flow of $D$ provides us with a boundary condition at the rightmost side of the new box, $y=1$. The numerical integration of this compactified version reveals that each initial condition converges towards a different fixed point, that is, to a single plateau (reminiscent of the plateaus 1 and 1' [see Fig.~\ref{fig_plateaus1a}] in the noncompactified version), different for each initial condition. This qualitative feature is not modified when the number of discretization points is increased and $m$ is varied and therefore highlights the fact that the previous stable fixed point, reached at large RG time and observed in Fig.~\ref{fig_plateaus1a} on the plateau 2, is a numerical artifact. However, the quantitative picture, that is, the precise positions of the plateaus, is modified when the number of discretization points is increased. We have not been able to obtain fully converged results by increasing the number of points in the grid which indicates that the behaviour of $D$ in the vicinity of $y\simeq1$ is not well captured by our numerical scheme in the compactified version.

The final remedy to these numerical hurdles is the following: going back to the noncompact formulation in terms of $\hat \phi$ and $\hat A$, we modify the way the derivatives of $\hat A$ are computed around $\hat \phi_{\rm max}$. Instead of using the ``five-point stencil'' method, we now fit the large-field region by a function $b \, \hat \phi^\gamma$ [where $\gamma$ is given by Eq.~(\ref{eq_largePhi})], and compute the derivatives at the boundary using this fitting function. This fit prevents the numerical drift that eventually leads the flow to leave the plateaus 1 or 1', and confirms that the fixed point $\eta_A^* \simeq 2.29$ is only a numerical artifact; see Fig.~\ref{fig_plateaus1b}.

From this numerical study, we conclude that the whole interval of fixed points with $\alpha \in [0,1[$ is stable, and the convergence to one of these fixed points is determined by the large-field behaviour of the initial condition. The importance of the initial condition due to the existence of this line of fixed points signals the breakdown of universality for this model, although a nontrivial anisotropy exponent $\alpha \neq 0$ is preserved.

\section{Conclusion}

To summarize, in all dimensions $d$ there exists a half-line of stable fixed points which correspond to a positive roughness exponent $\alpha$. In $d=2$ in particular, if one aims to study the effects of anisotropy, then only the fixed points for which $\zeta\geq1$ should be considered, which means that the line of fixed points shrinks to an interval in the case of the static noise ($\kappa=2$), or to a single (trivial) fixed point $\alpha=0$ for the isotropic noise ($\kappa=1$).

In the light of the results on this anisotropic model, it appears that the discussion about the origin of the scaling in erosional landscapes is not completely closed. However, some new elements are now available: anisotropy is indeed a relevant feature in this context, and should not be overlooked when modelizing erosion at short length scale. The nature of the noise is also a main characteristic and drastically modifies the scaling behaviour of the model, since it changes its universality class. As a remark, notice that within the NPRG formalism, the noise term could be studied for noninteger values of $\kappa$ between 1 and 2, therefore giving rise to a smaller range of accessible $\alpha$. The status of a noninteger value of $\kappa$ is not mathematically clear, but one can see it as an interpolation between the two meaningful values $\kappa=1$ (isotropic noise) and $\kappa=2$ (static noise).

Moreover, we believe that our results can give some insights for the great dispersion of the values of the roughness exponent $\alpha$ when looking at different field measurements: if this model is valid (or at least the fact that an interval of fixed points may be generic in more realistic erosion models), then the dispersion of the roughness exponent is a signature of this line of fixed points, each of them corresponding to a different value of the exponent due to the difference in the initial conditions, that is differences in the geological context in the case of real landscapes. Let us also emphasize the surprising yet interesting fact that even though this anisotropic model is rather simple (there is only one renormalized function), it yields a very nontrivial RG physics, functional in essence and displaying a line of fixed points. 

Finally, although this work was focused on the erosion of landscapes and on the topography itself, continuum models have also been devised and applied to river landscapes~\cite{giacometti1995,banavar1997}. Numerical studies stemming from these models have been carried out but they still lack a theoretical study. We believe that our framework could be applied successfully to these models, and will be subject to further work.

\section*{Acknowledgements}

We thank Jean-Marie Maillard for useful advice on differentially algebraic equations and for providing us with the exact solution of the fixed point equation given in Appendix~\ref{app_fixedPoint}, and we thank an anonymous referee for indicating Refs.~\cite{giacometti1995,banavar1997} to us. C.D. also thanks F\'elix Rose for useful discussions about the numerical scheme.

\appendix
\section{Derivation of the flow equations}
\label{app_flow}

In this appendix we derive the flow of the non-linear function $A_k(\phi)$, defined in Eq.~(\ref{eq_Ak}). Having in mind this definition, we use the Wetterich equation~(\ref{eq_Wetterich}) to deduce the following equality:
\begin{align}
\begin{split}
    &\partial_k \text{FT} \left(  \dfonc{\Gamma_k}{\tilde \phi (z)} \right) (\bm p) = -\frac{1}{2} \, \text{Tr} \int_{\bm k_1,\bm q_1,\bm q_2} \partial_k \mathcal{R}_k (\bm k_1)  \\
    & \cdot G_k(\!-\bm k_1,\!-\bm q_1; \phi ) \cdot \Gamma_{k,\tilde\psi}^{(3)}(\bm q_1,\bm q_2,\bm p) \cdot G_k(-\bm q_2,\bm k_1;\phi ) \, ,
\end{split}
\end{align}
where $\int_{\bm q}\equiv 1/(2\pi)^{d+1} \int_{q,\omega}\dd^{d-1}q_\bot \, \dd q_\parallel \, \dd \omega $, and $\Gamma_{k,\tilde\psi}^{(3)} \equiv \delta \Gamma^{(2)}_k/\delta \tilde\psi$ reads:
\begin{align}
\Gamma_{k,\tilde\psi}^{(3)}(\bm q_1,\bm q_2,\bm p)\!=\!\!\left( 
         \begin{array}{cc}\!\! p_{\parallel}^2 \, \text{TF}(A_k''(\phi))(\bm q_1+\bm q_2+\bm p) & 0 \\ 0 & 0 \end{array} 
    \right) \, .
\end{align}
Notice that we keep the same name for a function and its Fourier transform, such that a function $f(\bm q)$ has to be understood as the Fourier transform of $f(\bm x)$, and we recall the convention: $f(\bm q) = \int_{\bm x} f(\bm x) \ee^{-i(q x - \omega t)}$.

In order to get the flow of $A_k$, one now has to take the derivative of the previous expression with respect to $p_{\parallel}^2$, and then to evaluate it at $p=0$ and uniform field $\phi$. Since $\Gamma^{(3)}_k(\bm q_1,\bm q_2,\bm p) \propto p_{\parallel}^2$, the whole expression is proportional to $p_{\parallel}^2$ and the only non-vanishing term after the derivation and the evaluation at zero external momentum ($\bm p=0$) is the one obtained when deriving $\Gamma^{(3)}_k(\bm q_1,\bm q_2,\bm p)$ with respect to $p_{\parallel}^2$, and evaluating every other Fourier Transform at $\bm p=0$. This means that one can already perform the evaluation at constant field, which simplifies drastically the computation. One therefore gets:
\begin{align}
\begin{split}
     \partial_k A_k &= -\frac{1}{2} \, \text{Tr} \int_{\bm q_1} \partial_k \mathcal{R}_k (\bm q_1) \\
      &\cdot G_k(-\bm q_1;\phi) \cdot 
        \left( 
             \begin{array}{cc} A_k''(\phi) & 0 \\ 0 & 0 \end{array} 
        \right) 
    \cdot G_k(\bm q_1,\phi) \, ,
\end{split}
\end{align}
where the full propagator $G_k$ is now evaluated at uniform field and reads:
\begin{align}
G_k(\bm q;\phi) = \left( 
         \begin{array}{cc} \frac{2 W(\omega)}{P(q^2,\omega)P(q^2,-\omega)} & \frac{1}{P(q^2,-\omega)} \\ \frac{1}{P(q^2,\omega)} & 0 \end{array} 
    \right)  \, ,
\end{align}    
with $P(q^2,\omega)=R_k(q_\parallel^2,q_\bot^2)+q_\bot^2+q_\parallel^2 A_k'(\phi)+ i \omega $, and $W(\omega)=1$ for an isotropic noise, and $W(\omega)=\delta(\omega)$ for a static noise. After performing the matrix product and the trace, the integration over the frequencies is straightforward and yields for the flow of $A_k$:
\begin{align}
\begin{split}
    &\partial_k A_k = -\frac{(3\kappa-2) K_d}{2} \times \\
    &\int_{|q_\bot|=0}^{\infty} \int_{q_\parallel=-\infty}^\infty \frac{\partial_k R_k(q_\parallel^2,|q_\bot|^2) \, |q_\bot|^{d-2} A_k''(\phi) }{\left( R_k(q_\parallel^2,|q_\bot|^2)+|q_\bot|^2+q_\parallel^2 A_k'(\phi) \right)^{1+\kappa}}
    \label{eq_dkAdimensionfull}
\end{split}
\end{align}
where $\kappa=1$ for an isotropic noise, and $\kappa=2$ for a static noise, and where $K_d=(2^{d-1}\pi^{d/2}\Gamma(d/2))^{-1}=S_{d-1}/(2\pi)^d$ with $S_d$ the surface of the $d$-dimensional unit hypersphere. Notice that we have used the rotational invariance in the transverse direction to rewrite the integral over $q_\bot$ as an integral over its norm. Finally, one performs the change of variable $q_\parallel = \sqrt{y}\cos(\theta)$ and $q_\bot = \sqrt{y}\sin(\theta)$ with $y\in [0,\infty[$ and $\theta \in [0,\pi]$. If we furthermore chose the regulator $R_k$ to be a function of $y=q_\bot^2+q_\parallel^2$ only, we can write:
\begin{align}
    R_k(q_\parallel^2,|q_\bot|^2)= y k^2 r(y)  \, ,\label{eq_regu}
\end{align}
with $r(y)$ the usual momentum regulator, for example an exponential regulator:
\begin{align}
r(y) =  \frac{a}{\ee^y-1} \, ,
\label{eq_reguExp}
\end{align}
where $a$ is a free parameter. Finally, using the dimensionless variables as defined in Eq.~(\ref{eq_adimensionalisation}), the particular form of regulator~(\ref{eq_regu}) and  Eq.~(\ref{eq_dkAdimensionfull}) one finally gets the dynamical part of the flow, Eq.~(\ref{eq_flowAdyn}).

\section{Retrieving the one-loop perturbative results}
\label{app_perturbative}

To retrieve the perturbative results from~\cite{pastor-satorras1998,pastor-satorras1998a}, and from~\cite{antonov2017a}, we first evaluate the previous equations at the upper critical dimension $d_c$, which depends on the noise type: $d_c^{\, \text{stat}}=4$ for a static noise, and $d_c^{\, \text{iso}}=2$ for an isotropic noise. We define accordingly $\epsilon=d_c-d$. 

    \subsection{Pastor-Satorras and Rothman's results}
    
The equations derived in~\cite{pastor-satorras1998} are retrieved by performing a lowest-order expansion of the function $\hat A(\hat\phi)$:
\begin{align}
    \hat A(\hat\phi) = \hat \phi + \frac{\hat a_{3}}{3!} \hat\phi^3  \, , \label{eq_SatorExpansion}
\end{align}
where $\hat a_{1}\equiv 1$ by definition of the anomalous dimension $\eta_A$. Then, taking derivatives of the flow equation~(\ref{eq_flowA}), and evaluating them at $\hat \phi = 0$, one finds: 
\begin{align}
    \eta_A &= \frac{\epsilon}{2} + \frac{3\pi K_d}{8}\hat{a}_3 \label{eq_SatorEtaA} \, ,\\
    k\partial_k \hat{a}_3 &= - \epsilon \hat{a}_3+\frac{3\pi K_d}{2}\hat{a}_3^2 \, . \label{eq_SatorA3}
\end{align}
Notice that at first order in the $\epsilon$-expansion, the integrals of the dynamical part of the flow can be computed analytically at $d=d_c^{\, \text{stat}}$ or $d=d_c^{\, \text{iso}}$. Moreover, at the first-order in the $\epsilon$-expansion, one notices that the flow equations do not depend on the precise shape of the regulator $r(y)$. Finally, the definition of the term in front of the cubic term in $\hat A$, $\hat{a}_3$, differs from that of~\cite{pastor-satorras1998} and the relation between the two is $\hat a_{3}=2\lambda$. Their dimensionless parameter $\bar \lambda$ is also proportional to ours and we have the following relation between the two: $\hat a_3 = 2 (2\pi)^{d-1}/S_{d-1} \bar \lambda$ where $S_d$ is the surface area of a $d$-dimensional unit sphere. Up to these notation, and up to a factor $-1$ which comes from the fact their equations are derived for the real-space variable $l$, whereas ours are derived for the momentum $k$, Eq.~(\ref{eq_SatorA3}) is indeed equivalent to their Eq.~(6) in~\cite{pastor-satorras1998}. We also agree with their results for the roughness (and anisotropy) exponent, and the stable fixed point of Eqs.~(\ref{eq_SatorEtaA}) and (\ref{eq_SatorA3}) indeed yields:
\begin{align}
    \alpha \equiv (4\kappa-2d-\eta_A^*)/3 = \frac{5}{12} \epsilon \, .
\end{align}
We still emphasize that this result is not correct, even for $\epsilon \to 0$, because the expansion~(\ref{eq_SatorExpansion}) discards an infinity of equally relevant coupling constants and is thus not valid.

    \subsection{Antonov and Kakin's results}

Following~\cite{antonov2017a}, we set $\kappa=1$ (isotropic noise), $d_c=d_c^{\, \text{iso}}=2$ and we expand the function $\hat A(\hat\phi)$ as
\begin{align}
    \hat A(\hat\phi) = \hat\phi + \sum_{i=2}^\infty \frac{\hat a_{i}}{i!} \hat\phi^i \, .
\end{align}
Notice that $\hat A$ is not an odd function of $\hat\phi$. Again, taking derivatives of the flow equation~(\ref{eq_flowA}), and evaluating them at $\hat \phi = 0$, we are able to retrieve the equations derived in~\cite{antonov2017a}, except that we do not agree on their integration over the momenta. Indeed, in~\cite{antonov2017a}, the integration over the momenta $\int \dd \bm k$ seems to be performed as if $\bm k$ was isotropic, yielding a factor $S_d$ whereas we argued it should be a factor $S_{d-1}$. A factor $\pi$ coming from the integration over the angle $\theta$ is also missing. Up to this difference and notational discrepancies, our flow equations are in a one to one agreement with the $\beta$ functions of~\cite{antonov2017} (those of the first article~\cite{antonov2017a} involved a misprint in the $\beta_2$ function).

Notice also that contrary to what is stated in~\cite{antonov2017a}, taking $\hat a_{i}=0$ for all $i\neq3$ makes the RG equations of~\cite{antonov2017a} boil down to those of~\cite{pastor-satorras1998,pastor-satorras1998a} (up to the factor coming from the momentum integration discussed in the previous paragraph).

\section{Exact solution of the fixed point equation for $\eta_A^*=0$}
\label{app_fixedPoint}

In the special case of $\eta_A^*=0$ (which is not interesting for the physics since it means $\zeta=1/3<1$), the fixed point solution of the flow equation~(\ref{eq_flowAlitim}) can be solved exactly. Indeed, one can show that $\hat A'(\hat \phi)$ is a solution of the simple differential equation:
\begin{align}
4 \left(2 \hat \phi^2+5\right)^2 (\hat A')^3-\left(9 \hat A'+1\right)^2=0 \, ,
\end{align}
which can be solved exactly in terms of an integral over an algebraic integrand. In this special case, we therefore have a proof that a well-defined function exists on the whole real axis.

Moreover, this function is in fact also a solution of a \emph{linear} ordinary differential equation of order 4, on which  the study of the singularities can be performed. The main singularity lies at $\hat \phi^2 = -5/2$ and not on the real axis. Thus, at least in this case, the series expansion around $\hat\phi=0$ of the fixed point solution coincides with the fixed point solution, although it has a finite radius of convergence, $R=\sqrt{5/2}$. 

Although it is difficult to extrapolate this result to the physically interesting values of $\eta_A^*$, we have nonetheless checked that our numerical integration of the fixed point equation for $\eta_A^*=0$ matches this exact result.

\bibliography{biblio}

\begin{thebibliography}{66}%
\makeatletter
\providecommand \@ifxundefined [1]{%
 \@ifx{#1\undefined}
}%
\providecommand \@ifnum [1]{%
 \ifnum #1\expandafter \@firstoftwo
 \else \expandafter \@secondoftwo
 \fi
}%
\providecommand \@ifx [1]{%
 \ifx #1\expandafter \@firstoftwo
 \else \expandafter \@secondoftwo
 \fi
}%
\providecommand \natexlab [1]{#1}%
\providecommand \enquote  [1]{``#1''}%
\providecommand \bibnamefont  [1]{#1}%
\providecommand \bibfnamefont [1]{#1}%
\providecommand \citenamefont [1]{#1}%
\providecommand \href@noop [0]{\@secondoftwo}%
\providecommand \href [0]{\begingroup \@sanitize@url \@href}%
\providecommand \@href[1]{\@@startlink{#1}\@@href}%
\providecommand \@@href[1]{\endgroup#1\@@endlink}%
\providecommand \@sanitize@url [0]{\catcode `\\12\catcode `\$12\catcode
  `\&12\catcode `\#12\catcode `\^12\catcode `\_12\catcode `\%12\relax}%
\providecommand \@@startlink[1]{}%
\providecommand \@@endlink[0]{}%
\providecommand \url  [0]{\begingroup\@sanitize@url \@url }%
\providecommand \@url [1]{\endgroup\@href {#1}{\urlprefix }}%
\providecommand \urlprefix  [0]{URL }%
\providecommand \Eprint [0]{\href }%
\providecommand \doibase [0]{http://dx.doi.org/}%
\providecommand \selectlanguage [0]{\@gobble}%
\providecommand \bibinfo  [0]{\@secondoftwo}%
\providecommand \bibfield  [0]{\@secondoftwo}%
\providecommand \translation [1]{[#1]}%
\providecommand \BibitemOpen [0]{}%
\providecommand \bibitemStop [0]{}%
\providecommand \bibitemNoStop [0]{.\EOS\space}%
\providecommand \EOS [0]{\spacefactor3000\relax}%
\providecommand \BibitemShut  [1]{\csname bibitem#1\endcsname}%
\let\auto@bib@innerbib\@empty
\bibitem [{\citenamefont {Dodds}\ and\ \citenamefont
  {Rothman}(2000)}]{dodds2000}%
  \BibitemOpen
  \bibfield  {author} {\bibinfo {author} {\bibfnamefont {P.~S.}\ \bibnamefont
  {Dodds}}\ and\ \bibinfo {author} {\bibfnamefont {D.~H.}\ \bibnamefont
  {Rothman}},\ }\href {\doibase 10.1146/annurev.earth.28.1.571} {\bibfield
  {journal} {\bibinfo  {journal} {Annu. Rev. Earth Planet. Sci.}\ }\textbf
  {\bibinfo {volume} {28}},\ \bibinfo {pages} {571} (\bibinfo {year}
  {2000})}\BibitemShut {NoStop}%
\bibitem [{\citenamefont {Mandelbrot}(1982)}]{mandelbrot1982}%
  \BibitemOpen
  \bibfield  {author} {\bibinfo {author} {\bibfnamefont {B.}~\bibnamefont
  {Mandelbrot}},\ }\href@noop {} {\emph {\bibinfo {title} {The {{Fractal
  Geometry}} of {{Nature}}}}}\ (\bibinfo  {publisher} {{Freeman, San
  Francisco}},\ \bibinfo {year} {1982})\BibitemShut {NoStop}%
\bibitem [{\citenamefont {Horton}(1945)}]{horton1945}%
  \BibitemOpen
  \bibfield  {author} {\bibinfo {author} {\bibfnamefont {R.~E.}\ \bibnamefont
  {Horton}},\ }\href {\doibase 10.1130/0016-7606(1945)56[275:EDOSAT]2.0.CO;2}
  {\bibfield  {journal} {\bibinfo  {journal} {Geol. Soc. Am. Bull.}\ }\textbf
  {\bibinfo {volume} {56}},\ \bibinfo {pages} {275} (\bibinfo {year}
  {1945})}\BibitemShut {NoStop}%
\bibitem [{\citenamefont {Rodriguez-Iturbe}\ and\ \citenamefont
  {Rinaldo}(2001)}]{rodriguez-iturbe2001}%
  \BibitemOpen
  \bibfield  {author} {\bibinfo {author} {\bibfnamefont {I.}~\bibnamefont
  {Rodriguez-Iturbe}}\ and\ \bibinfo {author} {\bibfnamefont {A.}~\bibnamefont
  {Rinaldo}},\ }\href@noop {} {\emph {\bibinfo {title} {Fractal River Basins:
  {{Chance}} and Self-Organization}}}\ (\bibinfo  {publisher} {{Cambridge
  University Press, Cambridge}},\ \bibinfo {year} {2001})\BibitemShut {NoStop}%
\bibitem [{\citenamefont {Somfai}\ and\ \citenamefont
  {Sander}(1997)}]{somfai1997}%
  \BibitemOpen
  \bibfield  {author} {\bibinfo {author} {\bibfnamefont {E.}~\bibnamefont
  {Somfai}}\ and\ \bibinfo {author} {\bibfnamefont {L.~M.}\ \bibnamefont
  {Sander}},\ }\href {\doibase 10.1103/PhysRevE.56.R5} {\bibfield  {journal}
  {\bibinfo  {journal} {Phys. Rev. E}\ }\textbf {\bibinfo {volume} {56}},\
  \bibinfo {pages} {R5} (\bibinfo {year} {1997})}\BibitemShut {NoStop}%
\bibitem [{\citenamefont {Newman}\ and\ \citenamefont
  {Turcotte}(1990)}]{newman1990}%
  \BibitemOpen
  \bibfield  {author} {\bibinfo {author} {\bibfnamefont {W.~I.}\ \bibnamefont
  {Newman}}\ and\ \bibinfo {author} {\bibfnamefont {D.~L.}\ \bibnamefont
  {Turcotte}},\ }\href {http://gji.oxfordjournals.org/content/100/3/433.short}
  {\bibfield  {journal} {\bibinfo  {journal} {Geophys. J. Int.}\ }\textbf
  {\bibinfo {volume} {100}},\ \bibinfo {pages} {433} (\bibinfo {year}
  {1990})}\BibitemShut {NoStop}%
\bibitem [{\citenamefont {Mark}\ and\ \citenamefont
  {Aronson}(1984)}]{mark1984}%
  \BibitemOpen
  \bibfield  {author} {\bibinfo {author} {\bibfnamefont {D.~M.}\ \bibnamefont
  {Mark}}\ and\ \bibinfo {author} {\bibfnamefont {P.~B.}\ \bibnamefont
  {Aronson}},\ }\href {http://www.springerlink.com/index/U020V24627H68031.pdf}
  {\bibfield  {journal} {\bibinfo  {journal} {Math. Geol.}\ }\textbf {\bibinfo
  {volume} {16}},\ \bibinfo {pages} {671} (\bibinfo {year} {1984})}\BibitemShut
  {NoStop}%
\bibitem [{\citenamefont {Czir{\'o}k}\ \emph {et~al.}(1993)\citenamefont
  {Czir{\'o}k}, \citenamefont {Somfai},\ and\ \citenamefont
  {Vicsek}}]{czirok1993}%
  \BibitemOpen
  \bibfield  {author} {\bibinfo {author} {\bibfnamefont {A.}~\bibnamefont
  {Czir{\'o}k}}, \bibinfo {author} {\bibfnamefont {E.}~\bibnamefont {Somfai}},
  \ and\ \bibinfo {author} {\bibfnamefont {T.}~\bibnamefont {Vicsek}},\ }\href
  {https://journals.aps.org/prl/abstract/10.1103/PhysRevLett.71.2154}
  {\bibfield  {journal} {\bibinfo  {journal} {Phys. Rev. Lett.}\ }\textbf
  {\bibinfo {volume} {71}},\ \bibinfo {pages} {2154} (\bibinfo {year}
  {1993})}\BibitemShut {NoStop}%
\bibitem [{\citenamefont {Norton}\ and\ \citenamefont
  {Sorenson}(1989)}]{norton1989}%
  \BibitemOpen
  \bibfield  {author} {\bibinfo {author} {\bibfnamefont {D.}~\bibnamefont
  {Norton}}\ and\ \bibinfo {author} {\bibfnamefont {S.}~\bibnamefont
  {Sorenson}},\ }\href {http://www.springerlink.com/index/V40864828634410H.pdf}
  {\bibfield  {journal} {\bibinfo  {journal} {Pure Appl. Geophys.}\ }\textbf
  {\bibinfo {volume} {131}},\ \bibinfo {pages} {77} (\bibinfo {year}
  {1989})}\BibitemShut {NoStop}%
\bibitem [{\citenamefont {Ouchi}\ and\ \citenamefont
  {Matsushita}(1992)}]{ouchi1992}%
  \BibitemOpen
  \bibfield  {author} {\bibinfo {author} {\bibfnamefont {S.}~\bibnamefont
  {Ouchi}}\ and\ \bibinfo {author} {\bibfnamefont {M.}~\bibnamefont
  {Matsushita}},\ }\href
  {http://www.sciencedirect.com/science/article/pii/0169555X92900602}
  {\bibfield  {journal} {\bibinfo  {journal} {Geomorphology}\ }\textbf
  {\bibinfo {volume} {5}},\ \bibinfo {pages} {115} (\bibinfo {year}
  {1992})}\BibitemShut {NoStop}%
\bibitem [{\citenamefont {Matsushita}\ and\ \citenamefont
  {Ouchi}(1989)}]{matsushita1989}%
  \BibitemOpen
  \bibfield  {author} {\bibinfo {author} {\bibfnamefont {M.}~\bibnamefont
  {Matsushita}}\ and\ \bibinfo {author} {\bibfnamefont {S.}~\bibnamefont
  {Ouchi}},\ }\href {\doibase 10.1016/0167-2789(89)90201-7} {\bibfield
  {journal} {\bibinfo  {journal} {Phys. Nonlinear Phenom.}\ }\textbf {\bibinfo
  {volume} {38}},\ \bibinfo {pages} {246} (\bibinfo {year} {1989})}\BibitemShut
  {NoStop}%
\bibitem [{\citenamefont {Chase}(1992)}]{chase1992}%
  \BibitemOpen
  \bibfield  {author} {\bibinfo {author} {\bibfnamefont {C.~G.}\ \bibnamefont
  {Chase}},\ }\href
  {http://www.sciencedirect.com/science/article/pii/0169555X9290057U}
  {\bibfield  {journal} {\bibinfo  {journal} {Geomorphology}\ }\textbf
  {\bibinfo {volume} {5}},\ \bibinfo {pages} {39} (\bibinfo {year}
  {1992})}\BibitemShut {NoStop}%
\bibitem [{\citenamefont {Hasbargen}\ and\ \citenamefont
  {Paola}(2000)}]{hasbargen2000}%
  \BibitemOpen
  \bibfield  {author} {\bibinfo {author} {\bibfnamefont {L.~E.}\ \bibnamefont
  {Hasbargen}}\ and\ \bibinfo {author} {\bibfnamefont {C.}~\bibnamefont
  {Paola}},\ }\href {http://geology.gsapubs.org/content/28/12/1067.short}
  {\bibfield  {journal} {\bibinfo  {journal} {Geology}\ }\textbf {\bibinfo
  {volume} {28}},\ \bibinfo {pages} {1067} (\bibinfo {year}
  {2000})}\BibitemShut {NoStop}%
\bibitem [{\citenamefont {Paola}\ \emph {et~al.}(2009)\citenamefont {Paola},
  \citenamefont {Straub}, \citenamefont {Mohrig},\ and\ \citenamefont
  {Reinhardt}}]{paola2009}%
  \BibitemOpen
  \bibfield  {author} {\bibinfo {author} {\bibfnamefont {C.}~\bibnamefont
  {Paola}}, \bibinfo {author} {\bibfnamefont {K.}~\bibnamefont {Straub}},
  \bibinfo {author} {\bibfnamefont {D.}~\bibnamefont {Mohrig}}, \ and\ \bibinfo
  {author} {\bibfnamefont {L.}~\bibnamefont {Reinhardt}},\ }\href {\doibase
  10.1016/j.earscirev.2009.05.003} {\bibfield  {journal} {\bibinfo  {journal}
  {Earth-Sci. Rev.}\ }\textbf {\bibinfo {volume} {97}},\ \bibinfo {pages} {1}
  (\bibinfo {year} {2009})}\BibitemShut {NoStop}%
\bibitem [{\citenamefont {Kim}\ \emph {et~al.}(2000)\citenamefont {Kim},
  \citenamefont {Kim},\ and\ \citenamefont {Kim}}]{kim2000}%
  \BibitemOpen
  \bibfield  {author} {\bibinfo {author} {\bibfnamefont {H.-J.}\ \bibnamefont
  {Kim}}, \bibinfo {author} {\bibfnamefont {I.-M.}\ \bibnamefont {Kim}}, \ and\
  \bibinfo {author} {\bibfnamefont {J.~M.}\ \bibnamefont {Kim}},\ }\href
  {\doibase 10.1103/PhysRevE.62.3121} {\bibfield  {journal} {\bibinfo
  {journal} {Phys. Rev. E}\ }\textbf {\bibinfo {volume} {62}},\ \bibinfo
  {pages} {3121} (\bibinfo {year} {2000})}\BibitemShut {NoStop}%
\bibitem [{\citenamefont {Caldarelli}\ \emph {et~al.}(1997)\citenamefont
  {Caldarelli}, \citenamefont {Giacometti}, \citenamefont {Maritan},
  \citenamefont {Rodriguez-Iturbe},\ and\ \citenamefont
  {Rinaldo}}]{caldarelli1997}%
  \BibitemOpen
  \bibfield  {author} {\bibinfo {author} {\bibfnamefont {G.}~\bibnamefont
  {Caldarelli}}, \bibinfo {author} {\bibfnamefont {A.}~\bibnamefont
  {Giacometti}}, \bibinfo {author} {\bibfnamefont {A.}~\bibnamefont {Maritan}},
  \bibinfo {author} {\bibfnamefont {I.}~\bibnamefont {Rodriguez-Iturbe}}, \
  and\ \bibinfo {author} {\bibfnamefont {A.}~\bibnamefont {Rinaldo}},\ }\href
  {https://journals.aps.org/pre/abstract/10.1103/PhysRevE.55.R4865} {\bibfield
  {journal} {\bibinfo  {journal} {Phys. Rev. E}\ }\textbf {\bibinfo {volume}
  {55}},\ \bibinfo {pages} {R4865} (\bibinfo {year} {1997})}\BibitemShut
  {NoStop}%
\bibitem [{\citenamefont {Kalda}(2003)}]{kalda2003}%
  \BibitemOpen
  \bibfield  {author} {\bibinfo {author} {\bibfnamefont {J.}~\bibnamefont
  {Kalda}},\ }\href {\doibase 10.1103/PhysRevLett.90.118501} {\bibfield
  {journal} {\bibinfo  {journal} {Phys. Rev. Lett.}\ }\textbf {\bibinfo
  {volume} {90}},\ \bibinfo {pages} {118501} (\bibinfo {year}
  {2003})}\BibitemShut {NoStop}%
\bibitem [{\citenamefont {Kukal}(1990)}]{kukal1990}%
  \BibitemOpen
  \bibfield  {author} {\bibinfo {author} {\bibfnamefont {Z.}~\bibnamefont
  {Kukal}},\ }\href {\doibase 10.1016/0012-8252(90)90019-R} {\bibfield
  {journal} {\bibinfo  {journal} {Earth-Sci. Rev.}\ }\textbf {\bibinfo {volume}
  {28}},\ \bibinfo {pages} {10} (\bibinfo {year} {1990})}\BibitemShut {NoStop}%
\bibitem [{\citenamefont {Sornette}\ and\ \citenamefont
  {Zhang}(1993)}]{sornette1993}%
  \BibitemOpen
  \bibfield  {author} {\bibinfo {author} {\bibfnamefont {D.}~\bibnamefont
  {Sornette}}\ and\ \bibinfo {author} {\bibfnamefont {Y.-C.}\ \bibnamefont
  {Zhang}},\ }\href {http://gji.oxfordjournals.org/content/113/2/382.short}
  {\bibfield  {journal} {\bibinfo  {journal} {Geophys. J. Int.}\ }\textbf
  {\bibinfo {volume} {113}},\ \bibinfo {pages} {382} (\bibinfo {year}
  {1993})}\BibitemShut {NoStop}%
\bibitem [{\citenamefont {Kenyon}\ and\ \citenamefont
  {Turcotte}(1985)}]{kenyon1985}%
  \BibitemOpen
  \bibfield  {author} {\bibinfo {author} {\bibfnamefont {P.~M.}\ \bibnamefont
  {Kenyon}}\ and\ \bibinfo {author} {\bibfnamefont {D.~L.}\ \bibnamefont
  {Turcotte}},\ }\href
  {http://gsabulletin.gsapubs.org/content/96/11/1457.short} {\bibfield
  {journal} {\bibinfo  {journal} {Geol. Soc. Am. Bull.}\ }\textbf {\bibinfo
  {volume} {96}},\ \bibinfo {pages} {1457} (\bibinfo {year}
  {1985})}\BibitemShut {NoStop}%
\bibitem [{\citenamefont {Pelletier}(2007)}]{pelletier2007}%
  \BibitemOpen
  \bibfield  {author} {\bibinfo {author} {\bibfnamefont {J.~D.}\ \bibnamefont
  {Pelletier}},\ }\href {\doibase 10.1016/j.geomorph.2007.04.015} {\bibfield
  {journal} {\bibinfo  {journal} {Geomorphology}\ }\textbf {\bibinfo {volume}
  {91}},\ \bibinfo {pages} {291} (\bibinfo {year} {2007})}\BibitemShut
  {NoStop}%
\bibitem [{\citenamefont {Edwards}\ and\ \citenamefont
  {Wilkinson}(1982)}]{edwards1982}%
  \BibitemOpen
  \bibfield  {author} {\bibinfo {author} {\bibfnamefont {S.~F.}\ \bibnamefont
  {Edwards}}\ and\ \bibinfo {author} {\bibfnamefont {D.~R.}\ \bibnamefont
  {Wilkinson}},\ }\href {\doibase 10.1098/rspa.1982.0056} {\bibfield  {journal}
  {\bibinfo  {journal} {Proc. R. Soc. Math. Phys. Eng. Sci.}\ }\textbf
  {\bibinfo {volume} {381}},\ \bibinfo {pages} {17} (\bibinfo {year}
  {1982})}\BibitemShut {NoStop}%
\bibitem [{\citenamefont {Roering}\ \emph {et~al.}(1999)\citenamefont
  {Roering}, \citenamefont {Kirchner},\ and\ \citenamefont
  {Dietrich}}]{roering1999}%
  \BibitemOpen
  \bibfield  {author} {\bibinfo {author} {\bibfnamefont {J.~J.}\ \bibnamefont
  {Roering}}, \bibinfo {author} {\bibfnamefont {J.~W.}\ \bibnamefont
  {Kirchner}}, \ and\ \bibinfo {author} {\bibfnamefont {W.~E.}\ \bibnamefont
  {Dietrich}},\ }\href {\doibase 10.1029/1998WR900090} {\bibfield  {journal}
  {\bibinfo  {journal} {Water Resour. Res.}\ }\textbf {\bibinfo {volume}
  {35}},\ \bibinfo {pages} {853} (\bibinfo {year} {1999})}\BibitemShut
  {NoStop}%
\bibitem [{\citenamefont {Kardar}\ \emph {et~al.}(1986)\citenamefont {Kardar},
  \citenamefont {Parisi},\ and\ \citenamefont {Zhang}}]{kardar1986}%
  \BibitemOpen
  \bibfield  {author} {\bibinfo {author} {\bibfnamefont {M.}~\bibnamefont
  {Kardar}}, \bibinfo {author} {\bibfnamefont {G.}~\bibnamefont {Parisi}}, \
  and\ \bibinfo {author} {\bibfnamefont {Y.-C.}\ \bibnamefont {Zhang}},\ }\href
  {https://journals.aps.org/prl/abstract/10.1103/PhysRevLett.56.889} {\bibfield
   {journal} {\bibinfo  {journal} {Phys. Rev. Lett.}\ }\textbf {\bibinfo
  {volume} {56}},\ \bibinfo {pages} {889} (\bibinfo {year} {1986})}\BibitemShut
  {NoStop}%
\bibitem [{\citenamefont {Kloss}\ \emph {et~al.}(2012)\citenamefont {Kloss},
  \citenamefont {Canet},\ and\ \citenamefont {Wschebor}}]{kloss2012}%
  \BibitemOpen
  \bibfield  {author} {\bibinfo {author} {\bibfnamefont {T.}~\bibnamefont
  {Kloss}}, \bibinfo {author} {\bibfnamefont {L.}~\bibnamefont {Canet}}, \ and\
  \bibinfo {author} {\bibfnamefont {N.}~\bibnamefont {Wschebor}},\ }\href
  {\doibase 10.1103/PhysRevE.86.051124} {\bibfield  {journal} {\bibinfo
  {journal} {Phys. Rev. E}\ }\textbf {\bibinfo {volume} {86}},\ \bibinfo
  {pages} {051124} (\bibinfo {year} {2012})}\BibitemShut {NoStop}%
\bibitem [{\citenamefont {Kelling}\ and\ \citenamefont
  {{\'O}dor}(2011)}]{kelling2011}%
  \BibitemOpen
  \bibfield  {author} {\bibinfo {author} {\bibfnamefont {J.}~\bibnamefont
  {Kelling}}\ and\ \bibinfo {author} {\bibfnamefont {G.}~\bibnamefont
  {{\'O}dor}},\ }\href {\doibase 10.1103/PhysRevE.84.061150} {\bibfield
  {journal} {\bibinfo  {journal} {Phys. Rev. E}\ }\textbf {\bibinfo {volume}
  {84}},\ \bibinfo {pages} {061150} (\bibinfo {year} {2011})}\BibitemShut
  {NoStop}%
\bibitem [{\citenamefont {Pastor-Satorras}\ and\ \citenamefont
  {Rothman}(1998{\natexlab{a}})}]{pastor-satorras1998}%
  \BibitemOpen
  \bibfield  {author} {\bibinfo {author} {\bibfnamefont {R.}~\bibnamefont
  {Pastor-Satorras}}\ and\ \bibinfo {author} {\bibfnamefont {D.~H.}\
  \bibnamefont {Rothman}},\ }\href {\doibase 10.1103/PhysRevLett.80.4349}
  {\bibfield  {journal} {\bibinfo  {journal} {Phys. Rev. Lett.}\ }\textbf
  {\bibinfo {volume} {80}},\ \bibinfo {pages} {4349} (\bibinfo {year}
  {1998}{\natexlab{a}})}\BibitemShut {NoStop}%
\bibitem [{\citenamefont {Pastor-Satorras}\ and\ \citenamefont
  {Rothman}(1998{\natexlab{b}})}]{pastor-satorras1998a}%
  \BibitemOpen
  \bibfield  {author} {\bibinfo {author} {\bibfnamefont {R.}~\bibnamefont
  {Pastor-Satorras}}\ and\ \bibinfo {author} {\bibfnamefont {D.~H.}\
  \bibnamefont {Rothman}},\ }\href {\doibase
  10.1023/B:JOSS.0000033160.59155.c6} {\bibfield  {journal} {\bibinfo
  {journal} {J. Stat. Phys.}\ }\textbf {\bibinfo {volume} {93}},\ \bibinfo
  {pages} {477} (\bibinfo {year} {1998}{\natexlab{b}})}\BibitemShut {NoStop}%
\bibitem [{\citenamefont {Antonov}\ and\ \citenamefont
  {Kakin}(2017{\natexlab{a}})}]{antonov2017a}%
  \BibitemOpen
  \bibfield  {author} {\bibinfo {author} {\bibfnamefont {N.~V.}\ \bibnamefont
  {Antonov}}\ and\ \bibinfo {author} {\bibfnamefont {P.~I.}\ \bibnamefont
  {Kakin}},\ }\href {\doibase 10.1134/S0040577917020027} {\bibfield  {journal}
  {\bibinfo  {journal} {Theor. Math. Phys.}\ }\textbf {\bibinfo {volume}
  {190}},\ \bibinfo {pages} {193} (\bibinfo {year}
  {2017}{\natexlab{a}})}\BibitemShut {NoStop}%
\bibitem [{\citenamefont {Berges}\ \emph {et~al.}(2002)\citenamefont {Berges},
  \citenamefont {Tetradis},\ and\ \citenamefont {Wetterich}}]{berges2002}%
  \BibitemOpen
  \bibfield  {author} {\bibinfo {author} {\bibfnamefont {J.}~\bibnamefont
  {Berges}}, \bibinfo {author} {\bibfnamefont {N.}~\bibnamefont {Tetradis}}, \
  and\ \bibinfo {author} {\bibfnamefont {C.}~\bibnamefont {Wetterich}},\ }\href
  {\doibase 10.1016/S0370-1573(01)00098-9} {\bibfield  {journal} {\bibinfo
  {journal} {Phys. Rep.}\ }\textbf {\bibinfo {volume} {363}},\ \bibinfo {pages}
  {223} (\bibinfo {year} {2002})}\BibitemShut {NoStop}%
\bibitem [{\citenamefont {Gies}(2012)}]{gies2012}%
  \BibitemOpen
  \bibfield  {author} {\bibinfo {author} {\bibfnamefont {H.}~\bibnamefont
  {Gies}},\ }\href {\doibase 10.1007/978-3-642-27320-9_6} {\bibfield  {journal}
  {\bibinfo  {journal} {Lect. Notes Phys.}\ }\textbf {\bibinfo {volume}
  {852}},\ \bibinfo {pages} {287} (\bibinfo {year} {2012})}\BibitemShut
  {NoStop}%
\bibitem [{\citenamefont {Delamotte}(2012)}]{delamotte2012}%
  \BibitemOpen
  \bibfield  {author} {\bibinfo {author} {\bibfnamefont {B.}~\bibnamefont
  {Delamotte}},\ }\href {\doibase 10.1007/978-3-642-27320-9_2} {\bibfield
  {journal} {\bibinfo  {journal} {Lect. Notes Phys.}\ }\textbf {\bibinfo
  {volume} {852}},\ \bibinfo {pages} {49} (\bibinfo {year} {2012})}\BibitemShut
  {NoStop}%
\bibitem [{\citenamefont {Martin}\ \emph {et~al.}(1973)\citenamefont {Martin},
  \citenamefont {Siggia},\ and\ \citenamefont {Rose}}]{martin1973}%
  \BibitemOpen
  \bibfield  {author} {\bibinfo {author} {\bibfnamefont {P.~C.}\ \bibnamefont
  {Martin}}, \bibinfo {author} {\bibfnamefont {E.~D.}\ \bibnamefont {Siggia}},
  \ and\ \bibinfo {author} {\bibfnamefont {H.~A.}\ \bibnamefont {Rose}},\
  }\href {\doibase 10.1103/PhysRevA.8.423} {\bibfield  {journal} {\bibinfo
  {journal} {Phys. Rev. A}\ }\textbf {\bibinfo {volume} {8}},\ \bibinfo {pages}
  {423} (\bibinfo {year} {1973})}\BibitemShut {NoStop}%
\bibitem [{\citenamefont {Janssen}(1976)}]{janssen1976}%
  \BibitemOpen
  \bibfield  {author} {\bibinfo {author} {\bibfnamefont {H.-K.}\ \bibnamefont
  {Janssen}},\ }\href {\doibase 10.1007/BF01316547} {\bibfield  {journal}
  {\bibinfo  {journal} {Z. Phys. B}\ }\textbf {\bibinfo {volume} {23}},\
  \bibinfo {pages} {377} (\bibinfo {year} {1976})}\BibitemShut {NoStop}%
\bibitem [{\citenamefont {De~Dominicis}(1976)}]{dedominicis1976}%
  \BibitemOpen
  \bibfield  {author} {\bibinfo {author} {\bibfnamefont {C.}~\bibnamefont
  {De~Dominicis}},\ }\href {\doibase 10.1051/jphyscol:1976138} {\bibfield
  {journal} {\bibinfo  {journal} {J. Phys. Colloq.}\ }\textbf {\bibinfo
  {volume} {37}},\ \bibinfo {pages} {C1} (\bibinfo {year} {1976})}\BibitemShut
  {NoStop}%
\bibitem [{\citenamefont {Canet}\ and\ \citenamefont
  {Chat{\'e}}(2007)}]{canet2007}%
  \BibitemOpen
  \bibfield  {author} {\bibinfo {author} {\bibfnamefont {L.}~\bibnamefont
  {Canet}}\ and\ \bibinfo {author} {\bibfnamefont {H.}~\bibnamefont
  {Chat{\'e}}},\ }\href {\doibase 10.1088/1751-8113/40/9/002} {\bibfield
  {journal} {\bibinfo  {journal} {J. Phys. Math. Theor.}\ }\textbf {\bibinfo
  {volume} {40}},\ \bibinfo {pages} {1937} (\bibinfo {year}
  {2007})}\BibitemShut {NoStop}%
\bibitem [{\citenamefont {Canet}\ \emph
  {et~al.}(2011{\natexlab{a}})\citenamefont {Canet}, \citenamefont
  {Chat{\'e}},\ and\ \citenamefont {Delamotte}}]{canet2011a}%
  \BibitemOpen
  \bibfield  {author} {\bibinfo {author} {\bibfnamefont {L.}~\bibnamefont
  {Canet}}, \bibinfo {author} {\bibfnamefont {H.}~\bibnamefont {Chat{\'e}}}, \
  and\ \bibinfo {author} {\bibfnamefont {B.}~\bibnamefont {Delamotte}},\ }\href
  {\doibase 10.1088/1751-8113/44/49/495001} {\bibfield  {journal} {\bibinfo
  {journal} {J. Phys. Math. Theor.}\ }\textbf {\bibinfo {volume} {44}},\
  \bibinfo {pages} {495001} (\bibinfo {year} {2011}{\natexlab{a}})}\BibitemShut
  {NoStop}%
\bibitem [{\citenamefont {Duclut}\ and\ \citenamefont
  {Delamotte}(2017)}]{duclut2017}%
  \BibitemOpen
  \bibfield  {author} {\bibinfo {author} {\bibfnamefont {C.}~\bibnamefont
  {Duclut}}\ and\ \bibinfo {author} {\bibfnamefont {B.}~\bibnamefont
  {Delamotte}},\ }\href {\doibase 10.1103/PhysRevE.95.012107} {\bibfield
  {journal} {\bibinfo  {journal} {Phys. Rev. E}\ }\textbf {\bibinfo {volume}
  {95}},\ \bibinfo {pages} {012107} (\bibinfo {year} {2017})}\BibitemShut
  {NoStop}%
\bibitem [{\citenamefont {Wetterich}(1993)}]{wetterich1993}%
  \BibitemOpen
  \bibfield  {author} {\bibinfo {author} {\bibfnamefont {C.}~\bibnamefont
  {Wetterich}},\ }\href {\doibase 10.1016/0370-2693(93)90726-X} {\bibfield
  {journal} {\bibinfo  {journal} {Phys. Lett. B}\ }\textbf {\bibinfo {volume}
  {301}},\ \bibinfo {pages} {90} (\bibinfo {year} {1993})}\BibitemShut
  {NoStop}%
\bibitem [{\citenamefont {Morris}(1994)}]{morris1994}%
  \BibitemOpen
  \bibfield  {author} {\bibinfo {author} {\bibfnamefont {T.~R.}\ \bibnamefont
  {Morris}},\ }\href {\doibase 10.1142/S0217751X94000972} {\bibfield  {journal}
  {\bibinfo  {journal} {Int. J. Mod. Phys. A}\ }\textbf {\bibinfo {volume}
  {09}},\ \bibinfo {pages} {2411} (\bibinfo {year} {1994})}\BibitemShut
  {NoStop}%
\bibitem [{\citenamefont {Canet}\ \emph
  {et~al.}(2003{\natexlab{a}})\citenamefont {Canet}, \citenamefont {Delamotte},
  \citenamefont {Mouhanna},\ and\ \citenamefont {Vidal}}]{canet2003a}%
  \BibitemOpen
  \bibfield  {author} {\bibinfo {author} {\bibfnamefont {L.}~\bibnamefont
  {Canet}}, \bibinfo {author} {\bibfnamefont {B.}~\bibnamefont {Delamotte}},
  \bibinfo {author} {\bibfnamefont {D.}~\bibnamefont {Mouhanna}}, \ and\
  \bibinfo {author} {\bibfnamefont {J.}~\bibnamefont {Vidal}},\ }\href
  {\doibase 10.1103/PhysRevB.68.064421} {\bibfield  {journal} {\bibinfo
  {journal} {Phys. Rev. B}\ }\textbf {\bibinfo {volume} {68}},\ \bibinfo
  {pages} {064421} (\bibinfo {year} {2003}{\natexlab{a}})}\BibitemShut
  {NoStop}%
\bibitem [{\citenamefont {Canet}(2005)}]{canet2005a}%
  \BibitemOpen
  \bibfield  {author} {\bibinfo {author} {\bibfnamefont {L.}~\bibnamefont
  {Canet}},\ }\href {\doibase 10.1103/PhysRevB.71.012418} {\bibfield  {journal}
  {\bibinfo  {journal} {Phys. Rev. B}\ }\textbf {\bibinfo {volume} {71}},\
  \bibinfo {pages} {012418} (\bibinfo {year} {2005})}\BibitemShut {NoStop}%
\bibitem [{\citenamefont {Kloss}\ \emph {et~al.}(2014)\citenamefont {Kloss},
  \citenamefont {Canet}, \citenamefont {Delamotte},\ and\ \citenamefont
  {Wschebor}}]{kloss2014}%
  \BibitemOpen
  \bibfield  {author} {\bibinfo {author} {\bibfnamefont {T.}~\bibnamefont
  {Kloss}}, \bibinfo {author} {\bibfnamefont {L.}~\bibnamefont {Canet}},
  \bibinfo {author} {\bibfnamefont {B.}~\bibnamefont {Delamotte}}, \ and\
  \bibinfo {author} {\bibfnamefont {N.}~\bibnamefont {Wschebor}},\ }\href
  {\doibase 10.1103/PhysRevE.89.022108} {\bibfield  {journal} {\bibinfo
  {journal} {Phys. Rev. E}\ }\textbf {\bibinfo {volume} {89}},\ \bibinfo
  {pages} {022108} (\bibinfo {year} {2014})}\BibitemShut {NoStop}%
\bibitem [{\citenamefont {Canet}\ \emph
  {et~al.}(2011{\natexlab{b}})\citenamefont {Canet}, \citenamefont {Chat{\'e}},
  \citenamefont {Delamotte},\ and\ \citenamefont {Wschebor}}]{canet2011}%
  \BibitemOpen
  \bibfield  {author} {\bibinfo {author} {\bibfnamefont {L.}~\bibnamefont
  {Canet}}, \bibinfo {author} {\bibfnamefont {H.}~\bibnamefont {Chat{\'e}}},
  \bibinfo {author} {\bibfnamefont {B.}~\bibnamefont {Delamotte}}, \ and\
  \bibinfo {author} {\bibfnamefont {N.}~\bibnamefont {Wschebor}},\ }\href
  {\doibase 10.1103/PhysRevE.84.061128} {\bibfield  {journal} {\bibinfo
  {journal} {Phys. Rev. E}\ }\textbf {\bibinfo {volume} {84}},\ \bibinfo
  {pages} {061128} (\bibinfo {year} {2011}{\natexlab{b}})}\BibitemShut
  {NoStop}%
\bibitem [{\citenamefont {Canet}\ \emph {et~al.}(2012)\citenamefont {Canet},
  \citenamefont {Chat{\'e}}, \citenamefont {Delamotte},\ and\ \citenamefont
  {Wschebor}}]{canet2012}%
  \BibitemOpen
  \bibfield  {author} {\bibinfo {author} {\bibfnamefont {L.}~\bibnamefont
  {Canet}}, \bibinfo {author} {\bibfnamefont {H.}~\bibnamefont {Chat{\'e}}},
  \bibinfo {author} {\bibfnamefont {B.}~\bibnamefont {Delamotte}}, \ and\
  \bibinfo {author} {\bibfnamefont {N.}~\bibnamefont {Wschebor}},\ }\href
  {\doibase 10.1103/PhysRevE.86.019904} {\bibfield  {journal} {\bibinfo
  {journal} {Phys. Rev. E}\ }\textbf {\bibinfo {volume} {86}},\ \bibinfo
  {pages} {019904} (\bibinfo {year} {2012})}\BibitemShut {NoStop}%
\bibitem [{\citenamefont {Delamotte}\ and\ \citenamefont
  {Canet}(2004)}]{delamotte2004}%
  \BibitemOpen
  \bibfield  {author} {\bibinfo {author} {\bibfnamefont {B.}~\bibnamefont
  {Delamotte}}\ and\ \bibinfo {author} {\bibfnamefont {L.}~\bibnamefont
  {Canet}},\ }\href {https://arxiv.org/abs/cond-mat/0412205} {\bibfield
  {journal} {\bibinfo  {journal} {ArXiv Prepr Cond-Mat0412205}\ } (\bibinfo
  {year} {2004})}\BibitemShut {NoStop}%
\bibitem [{\citenamefont {Benitez}\ \emph {et~al.}(2008)\citenamefont
  {Benitez}, \citenamefont {M{\'e}ndez-Galain},\ and\ \citenamefont
  {Wschebor}}]{benitez2008}%
  \BibitemOpen
  \bibfield  {author} {\bibinfo {author} {\bibfnamefont {F.}~\bibnamefont
  {Benitez}}, \bibinfo {author} {\bibfnamefont {R.}~\bibnamefont
  {M{\'e}ndez-Galain}}, \ and\ \bibinfo {author} {\bibfnamefont
  {N.}~\bibnamefont {Wschebor}},\ }\href {\doibase 10.1103/PhysRevB.77.024431}
  {\bibfield  {journal} {\bibinfo  {journal} {Phys. Rev. B}\ }\textbf {\bibinfo
  {volume} {77}},\ \bibinfo {pages} {024431} (\bibinfo {year}
  {2008})}\BibitemShut {NoStop}%
\bibitem [{\citenamefont {Caffarel}\ \emph {et~al.}(2001)\citenamefont
  {Caffarel}, \citenamefont {Azaria}, \citenamefont {Delamotte},\ and\
  \citenamefont {Mouhanna}}]{caffarel2001}%
  \BibitemOpen
  \bibfield  {author} {\bibinfo {author} {\bibfnamefont {M.}~\bibnamefont
  {Caffarel}}, \bibinfo {author} {\bibfnamefont {P.}~\bibnamefont {Azaria}},
  \bibinfo {author} {\bibfnamefont {B.}~\bibnamefont {Delamotte}}, \ and\
  \bibinfo {author} {\bibfnamefont {D.}~\bibnamefont {Mouhanna}},\ }\href
  {\doibase 10.1103/PhysRevB.64.014412} {\bibfield  {journal} {\bibinfo
  {journal} {Phys. Rev. B}\ }\textbf {\bibinfo {volume} {64}},\ \bibinfo
  {pages} {014412} (\bibinfo {year} {2001})}\BibitemShut {NoStop}%
\bibitem [{\citenamefont {Holovatch}\ \emph {et~al.}(2004)\citenamefont
  {Holovatch}, \citenamefont {Ivaneyko},\ and\ \citenamefont
  {Delamotte}}]{holovatch2004}%
  \BibitemOpen
  \bibfield  {author} {\bibinfo {author} {\bibfnamefont {Y.}~\bibnamefont
  {Holovatch}}, \bibinfo {author} {\bibfnamefont {D.}~\bibnamefont {Ivaneyko}},
  \ and\ \bibinfo {author} {\bibfnamefont {B.}~\bibnamefont {Delamotte}},\
  }\href {http://iopscience.iop.org/article/10.1088/0305-4470/37/11/002/meta}
  {\bibfield  {journal} {\bibinfo  {journal} {J. Phys. Math. Gen.}\ }\textbf
  {\bibinfo {volume} {37}},\ \bibinfo {pages} {3569} (\bibinfo {year}
  {2004})}\BibitemShut {NoStop}%
\bibitem [{\citenamefont {Peles}\ \emph {et~al.}(2004)\citenamefont {Peles},
  \citenamefont {Southern}, \citenamefont {Delamotte}, \citenamefont
  {Mouhanna},\ and\ \citenamefont {Tissier}}]{peles2004}%
  \BibitemOpen
  \bibfield  {author} {\bibinfo {author} {\bibfnamefont {A.}~\bibnamefont
  {Peles}}, \bibinfo {author} {\bibfnamefont {B.~W.}\ \bibnamefont {Southern}},
  \bibinfo {author} {\bibfnamefont {B.}~\bibnamefont {Delamotte}}, \bibinfo
  {author} {\bibfnamefont {D.}~\bibnamefont {Mouhanna}}, \ and\ \bibinfo
  {author} {\bibfnamefont {M.}~\bibnamefont {Tissier}},\ }\href {\doibase
  10.1103/PhysRevB.69.220408} {\bibfield  {journal} {\bibinfo  {journal} {Phys.
  Rev. B}\ }\textbf {\bibinfo {volume} {69}},\ \bibinfo {pages} {220408}
  (\bibinfo {year} {2004})}\BibitemShut {NoStop}%
\bibitem [{\citenamefont {Delamotte}\ \emph {et~al.}(2004)\citenamefont
  {Delamotte}, \citenamefont {Mouhanna},\ and\ \citenamefont
  {Tissier}}]{delamotte2004a}%
  \BibitemOpen
  \bibfield  {author} {\bibinfo {author} {\bibfnamefont {B.}~\bibnamefont
  {Delamotte}}, \bibinfo {author} {\bibfnamefont {D.}~\bibnamefont {Mouhanna}},
  \ and\ \bibinfo {author} {\bibfnamefont {M.}~\bibnamefont {Tissier}},\ }\href
  {\doibase 10.1103/PhysRevB.69.134413} {\bibfield  {journal} {\bibinfo
  {journal} {Phys. Rev. B}\ }\textbf {\bibinfo {volume} {69}},\ \bibinfo
  {pages} {134413} (\bibinfo {year} {2004})}\BibitemShut {NoStop}%
\bibitem [{\citenamefont {Canet}\ \emph
  {et~al.}(2004{\natexlab{a}})\citenamefont {Canet}, \citenamefont {Delamotte},
  \citenamefont {Deloubri{\`e}re},\ and\ \citenamefont {Wschebor}}]{canet2004}%
  \BibitemOpen
  \bibfield  {author} {\bibinfo {author} {\bibfnamefont {L.}~\bibnamefont
  {Canet}}, \bibinfo {author} {\bibfnamefont {B.}~\bibnamefont {Delamotte}},
  \bibinfo {author} {\bibfnamefont {O.}~\bibnamefont {Deloubri{\`e}re}}, \ and\
  \bibinfo {author} {\bibfnamefont {N.}~\bibnamefont {Wschebor}},\ }\href
  {\doibase 10.1103/PhysRevLett.92.195703} {\bibfield  {journal} {\bibinfo
  {journal} {Phys. Rev. Lett.}\ }\textbf {\bibinfo {volume} {92}},\ \bibinfo
  {pages} {195703} (\bibinfo {year} {2004}{\natexlab{a}})}\BibitemShut
  {NoStop}%
\bibitem [{\citenamefont {Canet}\ \emph
  {et~al.}(2004{\natexlab{b}})\citenamefont {Canet}, \citenamefont
  {Chat{\'e}},\ and\ \citenamefont {Delamotte}}]{canet2004a}%
  \BibitemOpen
  \bibfield  {author} {\bibinfo {author} {\bibfnamefont {L.}~\bibnamefont
  {Canet}}, \bibinfo {author} {\bibfnamefont {H.}~\bibnamefont {Chat{\'e}}}, \
  and\ \bibinfo {author} {\bibfnamefont {B.}~\bibnamefont {Delamotte}},\ }\href
  {\doibase 10.1103/PhysRevLett.92.255703} {\bibfield  {journal} {\bibinfo
  {journal} {Phys. Rev. Lett.}\ }\textbf {\bibinfo {volume} {92}},\ \bibinfo
  {pages} {255703} (\bibinfo {year} {2004}{\natexlab{b}})}\BibitemShut
  {NoStop}%
\bibitem [{\citenamefont {Canet}\ \emph
  {et~al.}(2003{\natexlab{b}})\citenamefont {Canet}, \citenamefont {Delamotte},
  \citenamefont {Mouhanna},\ and\ \citenamefont {Vidal}}]{canet2003}%
  \BibitemOpen
  \bibfield  {author} {\bibinfo {author} {\bibfnamefont {L.}~\bibnamefont
  {Canet}}, \bibinfo {author} {\bibfnamefont {B.}~\bibnamefont {Delamotte}},
  \bibinfo {author} {\bibfnamefont {D.}~\bibnamefont {Mouhanna}}, \ and\
  \bibinfo {author} {\bibfnamefont {J.}~\bibnamefont {Vidal}},\ }\href
  {\doibase 10.1103/PhysRevD.67.065004} {\bibfield  {journal} {\bibinfo
  {journal} {Phys. Rev. D}\ }\textbf {\bibinfo {volume} {67}},\ \bibinfo
  {pages} {065004} (\bibinfo {year} {2003}{\natexlab{b}})}\BibitemShut
  {NoStop}%
\bibitem [{\citenamefont {Canet}\ \emph {et~al.}(2005)\citenamefont {Canet},
  \citenamefont {Chat{\'e}}, \citenamefont {Delamotte}, \citenamefont
  {Dornic},\ and\ \citenamefont {Mu{\~n}oz}}]{canet2005}%
  \BibitemOpen
  \bibfield  {author} {\bibinfo {author} {\bibfnamefont {L.}~\bibnamefont
  {Canet}}, \bibinfo {author} {\bibfnamefont {H.}~\bibnamefont {Chat{\'e}}},
  \bibinfo {author} {\bibfnamefont {B.}~\bibnamefont {Delamotte}}, \bibinfo
  {author} {\bibfnamefont {I.}~\bibnamefont {Dornic}}, \ and\ \bibinfo {author}
  {\bibfnamefont {M.~A.}\ \bibnamefont {Mu{\~n}oz}},\ }\href {\doibase
  10.1103/PhysRevLett.95.100601} {\bibfield  {journal} {\bibinfo  {journal}
  {Phys. Rev. Lett.}\ }\textbf {\bibinfo {volume} {95}},\ \bibinfo {pages}
  {100601} (\bibinfo {year} {2005})}\BibitemShut {NoStop}%
\bibitem [{\citenamefont {Tissier}\ and\ \citenamefont
  {Wschebor}(2010)}]{tissier2010}%
  \BibitemOpen
  \bibfield  {author} {\bibinfo {author} {\bibfnamefont {M.}~\bibnamefont
  {Tissier}}\ and\ \bibinfo {author} {\bibfnamefont {N.}~\bibnamefont
  {Wschebor}},\ }\href {\doibase 10.1103/PhysRevD.82.101701} {\bibfield
  {journal} {\bibinfo  {journal} {Phys. Rev. D}\ }\textbf {\bibinfo {volume}
  {82}},\ \bibinfo {pages} {101701} (\bibinfo {year} {2010})}\BibitemShut
  {NoStop}%
\bibitem [{\citenamefont {Tissier}\ and\ \citenamefont
  {Tarjus}(2008)}]{tissier2008}%
  \BibitemOpen
  \bibfield  {author} {\bibinfo {author} {\bibfnamefont {M.}~\bibnamefont
  {Tissier}}\ and\ \bibinfo {author} {\bibfnamefont {G.}~\bibnamefont
  {Tarjus}},\ }\href {\doibase 10.1103/PhysRevB.78.024204} {\bibfield
  {journal} {\bibinfo  {journal} {Phys. Rev. B}\ }\textbf {\bibinfo {volume}
  {78}},\ \bibinfo {pages} {024204} (\bibinfo {year} {2008})}\BibitemShut
  {NoStop}%
\bibitem [{\citenamefont {Tissier}\ and\ \citenamefont
  {Tarjus}(2012{\natexlab{a}})}]{tissier2012}%
  \BibitemOpen
  \bibfield  {author} {\bibinfo {author} {\bibfnamefont {M.}~\bibnamefont
  {Tissier}}\ and\ \bibinfo {author} {\bibfnamefont {G.}~\bibnamefont
  {Tarjus}},\ }\href {\doibase 10.1103/PhysRevB.85.104202} {\bibfield
  {journal} {\bibinfo  {journal} {Phys. Rev. B}\ }\textbf {\bibinfo {volume}
  {85}},\ \bibinfo {pages} {104202} (\bibinfo {year}
  {2012}{\natexlab{a}})}\BibitemShut {NoStop}%
\bibitem [{\citenamefont {Tissier}\ and\ \citenamefont
  {Tarjus}(2012{\natexlab{b}})}]{tissier2012a}%
  \BibitemOpen
  \bibfield  {author} {\bibinfo {author} {\bibfnamefont {M.}~\bibnamefont
  {Tissier}}\ and\ \bibinfo {author} {\bibfnamefont {G.}~\bibnamefont
  {Tarjus}},\ }\href {\doibase 10.1103/PhysRevB.85.104203} {\bibfield
  {journal} {\bibinfo  {journal} {Phys. Rev. B}\ }\textbf {\bibinfo {volume}
  {85}},\ \bibinfo {pages} {104203} (\bibinfo {year}
  {2012}{\natexlab{b}})}\BibitemShut {NoStop}%
\bibitem [{\citenamefont {Canet}\ \emph {et~al.}(2016)\citenamefont {Canet},
  \citenamefont {Delamotte},\ and\ \citenamefont {Wschebor}}]{canet2016}%
  \BibitemOpen
  \bibfield  {author} {\bibinfo {author} {\bibfnamefont {L.}~\bibnamefont
  {Canet}}, \bibinfo {author} {\bibfnamefont {B.}~\bibnamefont {Delamotte}}, \
  and\ \bibinfo {author} {\bibfnamefont {N.}~\bibnamefont {Wschebor}},\ }\href
  {\doibase 10.1103/PhysRevE.93.063101} {\bibfield  {journal} {\bibinfo
  {journal} {Phys. Rev. E}\ }\textbf {\bibinfo {volume} {93}},\ \bibinfo
  {pages} {063101} (\bibinfo {year} {2016})}\BibitemShut {NoStop}%
\bibitem [{\citenamefont {Antonov}\ and\ \citenamefont
  {Kakin}(2017{\natexlab{b}})}]{antonov2017}%
  \BibitemOpen
  \bibfield  {author} {\bibinfo {author} {\bibfnamefont {N.~V.}\ \bibnamefont
  {Antonov}}\ and\ \bibinfo {author} {\bibfnamefont {P.~I.}\ \bibnamefont
  {Kakin}},\ }\href {\doibase 10.1088/1751-8121/50/8/085002} {\bibfield
  {journal} {\bibinfo  {journal} {J. Phys. Math. Theor.}\ }\textbf {\bibinfo
  {volume} {50}},\ \bibinfo {pages} {085002} (\bibinfo {year}
  {2017}{\natexlab{b}})}\BibitemShut {NoStop}%
\bibitem [{\citenamefont {Bender}\ and\ \citenamefont
  {Orszag}(1978)}]{bender1978}%
  \BibitemOpen
  \bibfield  {author} {\bibinfo {author} {\bibfnamefont {C.~M.}\ \bibnamefont
  {Bender}}\ and\ \bibinfo {author} {\bibfnamefont {S.~A.}\ \bibnamefont
  {Orszag}},\ }\href@noop {} {\emph {\bibinfo {title} {Advanced Mathematical
  Methods for Scientists and Engineers}}}\ (\bibinfo  {publisher} {{Mac Graw
  Hill, New York City}},\ \bibinfo {year} {1978})\BibitemShut {NoStop}%
\bibitem [{\citenamefont {Bervillier}\ \emph
  {et~al.}(2008{\natexlab{a}})\citenamefont {Bervillier}, \citenamefont
  {Boisseau},\ and\ \citenamefont {Giacomini}}]{bervillier2008}%
  \BibitemOpen
  \bibfield  {author} {\bibinfo {author} {\bibfnamefont {C.}~\bibnamefont
  {Bervillier}}, \bibinfo {author} {\bibfnamefont {B.}~\bibnamefont
  {Boisseau}}, \ and\ \bibinfo {author} {\bibfnamefont {H.}~\bibnamefont
  {Giacomini}},\ }\href
  {http://www.sciencedirect.com/science/article/pii/S0550321307005354}
  {\bibfield  {journal} {\bibinfo  {journal} {Nucl. Phys. B}\ }\textbf
  {\bibinfo {volume} {789}},\ \bibinfo {pages} {525} (\bibinfo {year}
  {2008}{\natexlab{a}})}\BibitemShut {NoStop}%
\bibitem [{\citenamefont {Bervillier}\ \emph
  {et~al.}(2008{\natexlab{b}})\citenamefont {Bervillier}, \citenamefont
  {Boisseau},\ and\ \citenamefont {Giacomini}}]{bervillier2008a}%
  \BibitemOpen
  \bibfield  {author} {\bibinfo {author} {\bibfnamefont {C.}~\bibnamefont
  {Bervillier}}, \bibinfo {author} {\bibfnamefont {B.}~\bibnamefont
  {Boisseau}}, \ and\ \bibinfo {author} {\bibfnamefont {H.}~\bibnamefont
  {Giacomini}},\ }\href
  {http://www.sciencedirect.com/science/article/pii/S0550321308001430}
  {\bibfield  {journal} {\bibinfo  {journal} {Nucl. Phys. B}\ }\textbf
  {\bibinfo {volume} {801}},\ \bibinfo {pages} {296} (\bibinfo {year}
  {2008}{\natexlab{b}})}\BibitemShut {NoStop}%
\bibitem [{\citenamefont {Giacometti}\ \emph {et~al.}(1995)\citenamefont
  {Giacometti}, \citenamefont {Maritan},\ and\ \citenamefont
  {Banavar}}]{giacometti1995}%
  \BibitemOpen
  \bibfield  {author} {\bibinfo {author} {\bibfnamefont {A.}~\bibnamefont
  {Giacometti}}, \bibinfo {author} {\bibfnamefont {A.}~\bibnamefont {Maritan}},
  \ and\ \bibinfo {author} {\bibfnamefont {J.~R.}\ \bibnamefont {Banavar}},\
  }\href {https://journals.aps.org/prl/abstract/10.1103/PhysRevLett.75.577}
  {\bibfield  {journal} {\bibinfo  {journal} {Phys. Rev. Lett.}\ }\textbf
  {\bibinfo {volume} {75}},\ \bibinfo {pages} {577} (\bibinfo {year}
  {1995})}\BibitemShut {NoStop}%
\bibitem [{\citenamefont {Banavar}\ \emph {et~al.}(1997)\citenamefont
  {Banavar}, \citenamefont {Colaiori}, \citenamefont {Flammini}, \citenamefont
  {Giacometti}, \citenamefont {Maritan},\ and\ \citenamefont
  {Rinaldo}}]{banavar1997}%
  \BibitemOpen
  \bibfield  {author} {\bibinfo {author} {\bibfnamefont {J.~R.}\ \bibnamefont
  {Banavar}}, \bibinfo {author} {\bibfnamefont {F.}~\bibnamefont {Colaiori}},
  \bibinfo {author} {\bibfnamefont {A.}~\bibnamefont {Flammini}}, \bibinfo
  {author} {\bibfnamefont {A.}~\bibnamefont {Giacometti}}, \bibinfo {author}
  {\bibfnamefont {A.}~\bibnamefont {Maritan}}, \ and\ \bibinfo {author}
  {\bibfnamefont {A.}~\bibnamefont {Rinaldo}},\ }\href
  {https://journals.aps.org/prl/abstract/10.1103/PhysRevLett.78.4522}
  {\bibfield  {journal} {\bibinfo  {journal} {Phys. Rev. Lett.}\ }\textbf
  {\bibinfo {volume} {78}},\ \bibinfo {pages} {4522} (\bibinfo {year}
  {1997})}\BibitemShut {NoStop}%
\end{thebibliography}%

\end{document}